\pdfoutput=1
\documentstyle[11pt,graphicx,color,cancel,cite,ulem,framed,hyperref,url,subfigure,amsmath,amssymb]{article}

\def\issue(#1,#2,#3){{\bf #1}, #2 (#3)}

\catcode`\@=11
\def\lsim{\mathrel{\mathpalette\@versim<}}
\def\gsim{\mathrel{\mathpalette\@versim>}}
\def\@versim#1#2{\vcenter{\offinterlineskip
\ialign{$\m@th#1\hfil##\hfil$\crcr#2\crcr\sim\crcr } }}
\catcode`\@=12

\catcode`@=12
\hypersetup{
   bookmarks=true,         % show bookmarks bar?
   unicode=false,          % non-Latin characters in AcrobatÍs bookmarks
   pdftoolbar=true,        % show AcrobatÍs toolbar?
   pdfmenubar=true,        % show AcrobatÍs menu?
   pdffitwindow=false,     % window fit to page when opened
   pdfstartview={FitH},    % fits the width of the page to the window
   pdftitle={My title},    % title
   pdfauthor={Author},     % author
   pdfsubject={Subject},   % subject of the document
   pdfcreator={Creator},   % creator of the document
   pdfproducer={Producer}, % producer of the document
   pdfkeywords={keyword1} {key2} {key3}, % list of keywords
   pdfnewwindow=true,      % links in new window
   colorlinks=true,       % false: boxed links; true: colored links
   linkcolor=blue,        % color of internal links
   citecolor=red,         % color of links to bibliography
   filecolor=magenta,      % color of file links
   urlcolor=cyan,           % color of external links
   linktocpage = true,
   }

\def\beq {\begin{equation}}
\def\eeq {\end{equation}}
\def\bi {\begin{itemize}}
\def\ei {\end{itemize}}
\def\bea {\begin{eqnarray}}
\def\eea {\end{eqnarray}}

\def \MET{\rm E{\!\!\!/}_T}
\def \met{\rm E{\!\!\!/}_T}

\newcommand{\br}{\begin{eqnarray}}
\newcommand{\er}{\end{eqnarray}}
\newcommand{\be}{\begin{equation}}
\newcommand{\ee}{\end{equation}}

\def\lum             {{\cal L}}
\newcommand{\ifb} {\rm {fb}^{-1}}

\def\chisq{\chi^2}

\begin{document}

\begin{flushright}
{TIFR/TH/15-10, HRI-RECAPP-2015-006}
\end{flushright}

\begin{center}

{\large \bf Status of MSSM Higgs Sector using Global Analysis and Direct Search Bounds, and 
Future Prospects at the HL-LHC} \\    
%or\\
%\tcb{\large \bf Status of Non-decoupling MSSM Higgs sector using Global analysis and Direct search bounds, and 
%Future prospects at the HL-LHC}    

\vskip 0.5cm
Biplob Bhattacherjee$^{a}$\footnote{biplob@cts.iisc.ernet.in},
Amit Chakraborty$^{b}$\footnote{amit@theory.tifr.res.in}, 
and Arghya Choudhury$^{c}$\footnote{arghyachoudhury@hri.res.in}
\vskip 0.3cm

{$^a$  Centre for High Energy Physics, \\ 
Indian Institute of Science, Bangalore 560012, India}

\vskip 0.1cm
{$^b$  Department of Theoretical Physics,  
Tata Institute of Fundamental Research,\\
1, Homi Bhabha Road, Mumbai 400005, India}

\vskip 0.1cm
{$^c$ Regional Centre for Accelerator-based Particle Physics,\\
Harish-Chandra Research Institute, Jhunsi, Allahabad - 211019, India}

\end{center}

%\pacs{}

\vskip 0.3cm 
\begin{abstract}

  In this paper, we search for the regions of the phenomenological minimal
supersymmetric standard model (pMSSM) parameter space where one can expect to
have moderate Higgs mixing angle ($\alpha$) with relatively light (up to 600
GeV) additional Higgses after satisfying the current LHC data. We perform a
global fit analysis using most updated data (till December 2014) from the LHC
and Tevatron experiments. The constraints coming from the precision
measurements of the rare b-decays $B_s \to \mu^+ \mu^-$ and $b\to s \gamma$ are
also considered. We find that low $M_{A}$ $(\lesssim 350)$ and high $\tan\beta$
$(\gtrsim 25)$ regions are disfavored by the combined effect of the global
analysis and flavour data. However, regions with Higgs mixing angle $\alpha
\sim$ 0.1 - 0.8 are still allowed by the current data. We then study the
existing direct search bounds on the heavy scalar/pseudoscalar ($\rm H/A$) and
charged Higgs boson ($\rm H^\pm$) masses and branchings at the LHC. It has been found
that regions with low to moderate values of $\tan\beta$ with light additional
Higgses (mass $\le$ 600 GeV) are unconstrained by the data, while the regions
with $\tan\beta >$ 20 are excluded considering the direct search bounds by the
LHC-8 data. The possibility to probe the region with $\tan\beta$ $\le$ 20 
at the high luminosity run of LHC are also discussed, giving special 
attention to the $\rm H \to hh$, $\rm H/{A} \to t \bar t$ and 
$\rm H/{A} \to \tau^{+}\tau^{-}$ decay modes. 

% It has been found that the regions of parameter space up to $\tan\beta$ $\sim$ 8 with 
% low to moderate values of $\rm M_A$ are expected to be probed  at the high luminosity 
% (3000 $\rm fb^{-1}$) run of LHC-14.} 

\end{abstract}

%%%%%%%%%%%%%%%%%%%%%%%%%%%%%%%%%%%%%%%%%%%%%%%
\newpage
\setcounter{footnote}{0}
%%%%%%%%%%%%%%%%%%%%%% Table of content %%%%%%%
%\hypertarget{toc}{}
%\small
\hrule
\tableofcontents
\vskip 1.0cm
\hrule
%%%%%%%%%%%%%%%%%%%%%%%%%%%%%%%%%%%%%%%%%%%%%%%
%%%%%%%%%%%%%%%%%%%%%%%%%%%%%%%%%%%%%%%%%%%%%%%
%\newpage 

\section{Introduction}

The ATLAS and CMS collaborations have now confirmed the discovery 
of a charge neutral spin-0 particle with mass close to 125 GeV 
using the 7+8 TeV data at the LHC \cite{atlas_125,cms_125}. The CDF 
and $D\emptyset$ collaborations at the Tevatron have also reported the evidence 
of a SM-like Higgs boson with mass around 125 GeV 
\cite{Aaltonen:2013kxa} with Higgs boson decaying to $b\bar b$. 
Measurement of it's various couplings with the Standard Model (SM) 
particles so far seems to be consistent with the predictions 
of the SM Higgs boson. The primary goal of the run-II of the 
LHC with increased center-of-mass energy and 
integrated luminosity would be the more detailed study of 
various properties of the observed Higgs boson. The production 
modes of the observed Higgs boson that are analyzed at the LHC 
are the gluon-gluon fusion (ggF), vector boson fusion (VBF), 
associated production with a $\rm W$ or $\rm Z$ bosons, 
and associated production with a top-antitop pair, while 
the decay modes of the Higgs boson that are 
studied by the ATLAS and CMS 
collaborations\footnote{Recently, search for the SM-like 
Higgs boson decaying 
to $\mu^{+}\mu^{-}$ and $e^{+}e^{-}$ are also performed by both 
the ATLAS and CMS collaborations with 7+8 TeV LHC 
data~\cite{Aad:2014xva, Khachatryan:2014aep}.} are 
$\gamma\gamma$ \cite{Aad:2014eha,Khachatryan:2014ira}, 
$\rm WW$ \cite{ATLAS:2014aga,Chatrchyan:2013iaa}, $\rm ZZ$ \cite{
Aad:2014eva,Chatrchyan:2013mxa}, $b\bar b$ \cite{
Aad:2014xzb,Chatrchyan:2013zna} and 
$\tau^+\tau^-$ \cite{atlastautau,Chatrchyan:2014nva}. Note that, 
the di-photon and $\rm ZZ$ modes have been used to precisely 
measure the mass of the observed Higgs boson 
\cite{Aad:2014aba,Khachatryan:2014jba}. Even though various 
coupling measurements at the LHC 
have already ruled out large deviations from the 
SM expectations, still the presence 
of uncertainties in coupling measurements do 
not rule out the possibility of having non-standard 
couplings of the observed Higgs boson. In fact, any 
small but statistically significant deviation from 
the SM expectations can be thought of as the first 
indication of new physics.

Supersymmetry (SUSY) \cite{susy,Martin:1997ns,Djouadi:2005gj} 
has been one of the most popular extensions of the SM, 
however a SUSY signature is yet to be observed at the LHC. 
Non-observation of sparticles at the LHC have placed 
severe constraints on various superparticle 
masses and couplings \cite{cmstwiki,atlastwiki}. The Higgs 
sector of the CP conserving minimal supersymmetric 
standard model (MSSM) contains two CP-even neutral 
Higgs bosons $\rm h$ and $\rm H$,
one CP-odd neutral Higgs boson $\rm A$,
and two charged Higgs bosons $\rm H^\pm$. One can identify 
the observed 125 GeV Higgs boson with the lightest MSSM 
Higgs boson $h$. At the tree level, the Higgs sector of MSSM is 
described by two parameters: pseudoscalar mass $M_A$ and 
$\tan\beta$, where $\tan\beta$ is the ratio of the vacuum 
expectation values (vevs) of the two Higgs doublets $H_u$ and 
$H_d$. The mixing angle $\alpha$ between 
the neutral components of the two Higgs doublets can be 
determined in terms of $M_A$ and $\tan\beta$ at the tree 
level. However, 
radiative corrections to the Higgs boson mass matrix 
involving various SUSY parameters can modify the 
tree level value of $\alpha$ significantly. Moreover, 
the couplings of the lightest Higgs boson $h$ with the 
SM particles depend on $\alpha$ and $\beta$, and thus a 
global fit analysis considering various Higgs 
coupling measurements at the LHC and Tevatron experiments 
can, in principle, constrain the MSSM parameter space. 

Soon after the discovery of the 125 GeV Higgs boson, 
several analyses have been performed, for example, in 
the context of MSSM \cite{Ellis:2012rx,Ellis:2012hz,Klute:2012pu,Belanger:2012gc,
Djouadi:2013qya,Ellis:2013lra,Giardino:2013bma,
Cheung:2013kla,Corbett:2012ja,Plehn:2012iz,Cheung:2015uia,deVries:2015hva,
deBlas:2014ula,Cheung:2014oaa,Cacciapaglia:2012wb,Endo:2015oia}, general 2-Higgs doublet 
models \cite{Cheung:2013rva,Belanger:2013xza,Celis:2013rcs, Altmannshofer:2012ar,
Bai:2012ex,Chang:2012ve,Cheon:2012rh}, next-to minimal supersymmetric 
standard model (NMSSM) \cite{Ellwanger:2014dfa}, effective theory framework 
\cite{Carmi:2012yp,Banerjee:2012xc,Chang:2013cia,Ellis:2014jta,Ellis:2014dva,
Banerjee:2013apa,Espinosa:2012in}.  
%and some other possible extensions of SM \cite{Dey:2013cqa,Banerjee:2013hxa}. 
However, in the last few months, both the ATLAS and CMS collaborations have updated 
some of their analyses on Higgs signal strength measurements. For example, 
ATLAS has updated their results for Higgs decaying to 
$\gamma\gamma$ \cite{Aad:2014eha}, $\rm WW$ \cite{ATLAS:2014aga}, 
$\rm ZZ$ \cite{Aad:2014eva}, $b\bar b$ \cite{Aad:2014xzb} channels, while 
CMS has published their new result for Higgs decaying to $\gamma\gamma$ 
\cite{Khachatryan:2014ira}. In most of their analyses, the measure 
of uncertainty associated to various Higgs couplings have been now 
reduced by a sizable amount. A global analysis with the most 
updated (up to December 2014) results on various Higgs 
coupling measurements at the LHC and Tevatron would be extremely 
useful to probe the MSSM parameter space.

In order to probe the Higgs sector of the MSSM, the discovery or 
exclusion of additional Higgs boson is extremely crucial 
at the LHC. In addition to several extensive studies of 
the sparticles of the MSSM, both the ATLAS and 
CMS collaborations have also performed dedicated searches 
of these additional Higgses in various possible final 
state signatures at the LHC. Searches for the heavy 
Higgses (both $\rm H$ and $\rm A$) are 
performed at the LHC when they are produced via 
the gluon-gluon fusion and b-associated production processes 
with their decay to a pair 
of $\rm \tau$s \cite{Aad:2014vgg,Khachatryan:2014wca}. 
The ATLAS and CMS collaborations have also searched for the 
heavy resonances with the following decay modes: 
$H\to \gamma\gamma$ \cite{CMS:2014onr}, 
$H\to hh\to b\bar b \gamma\gamma$ \cite{CMS:2014ipa}
and $H\to hh\to b\bar b b \bar b$ \cite{CMS:2014eda}. 
Heavy SM-like Higgs bosons decaying to pair of $\rm W$ bosons with 
final state consisting of one lepton, two jets and missing transverse 
energy has been studied by the CMS collaboration \cite{CMS:2012bea}. 
Heavy pseudoscalar Higgs boson (A) decaying to a $\rm Z$ boson and a 
light Higgs boson $\rm h$ has also been searched by both the ATLAS and 
CMS collaborations at 19.7 $\rm fb^{-1}$ luminosity
at the 8 TeV run of LHC \cite{Aad:2015wra,CMS:2014yra}. In 
addition, search for the charged Higgs bosons ($\rm H^\pm$) decaying 
to $\tau\nu_{\tau}$, $c\bar s$ and $t\bar b$ are also performed 
at the LHC when $\rm H^{\pm}$ are produced via 
$t\bar t$ or in association with top quarks 
\cite{Aad:2014kga,CMS:2014pea}. So far, the ATLAS and CMS data 
have not revealed any clear signature of the signal and thus they put 
model-independent 95\% C.L. upper limits on the production 
cross section times branching ratios for different production 
processes and decay modes.

In the MSSM, the couplings of $h$ with the SM electro-weak gauge 
bosons are proportional to $\sin(\beta - \alpha)$. Now, the 
measured values of the couplings of the observed 125 GeV Higgs 
boson with the SM $W/Z$ bosons are quite consistent with 
the SM expectations, which thus restricts the value 
of $\sin(\beta - \alpha)$ near to unity. Besides, 
non observation of additional Higgses ($H$, $A$ and $H^{\pm}$) 
at the LHC implies that their masses are possibly well above the 
electro-weak scale. One can satisfy both the 
above-mentioned observations in the MSSM by 
choosing $\alpha \sim 0$ and $\beta \sim \frac{\pi}{2}$ 
and/or $M_{A} >> M_{Z}$, which is generally 
known as the decoupling limit of the MSSM \cite{Djouadi:2005gj}. 
In this paper, however, we would like to address the following 
question: Is there any  
%question: Does there exist any  
MSSM parameter space still allowed 
by the current LHC data where one can expect to have moderate Higgs 
mixing angle ($\alpha$) with relatively light 
(say, hundred to few hundred GeV) additional Higgses ? In other words,  
we are looking for a feasible MSSM parameter space where 
the non-negligible Higgs mixing still exists, and 
simultaneously we have light additional 
scalar particles which can be probed at the run-II 
of LHC. Our strategy can summarized as follows:

\bi
\item We first perform a global $\chi^{2}$ analysis by scanning the
relevant parameters of the MSSM Higgs sector and incorporating the 
updated Higgs signal strengths from the ATLAS, CMS and Tevatron
experiments. The constraints coming from the precision 
measurements of rare decays like $B_s \to \mu^+ \mu^-$ 
and $b\to s \gamma$ are also considered.

\item We then impose the LHC direct search bounds on the 
heavy scalar/pseudoscalar Higgs ($\rm H/A$) and 
charged Higgs boson ($\rm H^\pm$) masses and various branchings 
in the allowed parameter space.
\item Finally, we study the possibility to probe the remaining 
parameter space in the high luminosity run of LHC. 
\ei

In Sec.~\ref{sec2}, we discuss the detailed prescription 
of our global fit analysis, followed by a brief outline of our 
parameter space scan. The favorable parameter space obtained 
after the global analysis is discussed in Sec.~\ref{sec21}. 
In Sec.~\ref{sec3}, we present the 
current limits on production 
cross section times branching ratios obtained from the direct search of both 
neutral and charged Higgs bosons at the LHC. Future limits for 
the high luminosity run of LHC are discussed in Sec.~\ref{sec4}. 
Finally in Sec.~\ref{sec5} 
we summarize our results.

%%=================================================================================
%------------------- Sec. 2 starts -------------------------------------------------- 
%%=================================================================================

\section{Global analysis and available pMSSM parameter space}
\label{sec2}

In this section, we discuss the details of our global 
fit analysis. Current bounds on various Higgs 
coupling measurements by the ATLAS and CMS collaborations at 
the LHC, and also by the CDF and $D\emptyset$ collaborations at the 
Tevatron experiment are considered as the inputs to our 
global analysis. We also consider two important flavour physics 
constraints, namely ${\rm Br}(b \to s \gamma)$ and 
${\rm Br}(B_{s} \to \mu^+ \mu^-)$  \cite{Amhis:2014hma}. Finally, we discuss 
the features of the available parameter space satisfying 
the updated Higgs and flavour physics data.   
 
\subsection{Experimental inputs and Global fit analysis}
\label{sec:globalfit}

Both the ATLAS and CMS collaborations have published 
the results of the 125 GeV Higgs boson searches combining 
the 7+8 TeV data at the end of 8 TeV run of LHC 
\cite{Aad:2014eha,ATLAS:2014aga,Aad:2014eva,Aad:2014xzb,
atlastautau,Aad:2014aba,Khachatryan:2014ira,
Chatrchyan:2013iaa,Chatrchyan:2013mxa,Chatrchyan:2013zna,
Chatrchyan:2014nva,Khachatryan:2014jba}. Besides, the 
results of the SM Higgs boson search by the CDF and $D\emptyset$ collaborations at 
the Tevatron experiment are also available in the literature 
\cite{Aaltonen:2013kxa,Junk:2014yea}. In our global analysis, 
we consider the most updated Higgs data obtained from 
the LHC and Tevatron experiments. The Higgs bosons are produced at the LHC 
mainly via the gluon-gluon fusion (ggF) process. However, there 
exist other sub-dominant production mechanisms e.g., vector 
boson fusion (VBF), associated production with a $W/Z$ boson (Vh), 
associated production with a pair of top quarks ($t \bar t h$). 
The decay modes of the Higgs boson which are analyzed by the 
ATLAS and CMS collaborations are  
$h \to \gamma \gamma$ \cite{Aad:2014eha,Khachatryan:2014ira}, 
$h \to WW^*$ \cite{ATLAS:2014aga,Chatrchyan:2013iaa}, 
$h \to ZZ^*$ \cite{Aad:2014eva,Chatrchyan:2013mxa}, 
$h \to b\bar b$ \cite{Aad:2014xzb,Chatrchyan:2013zna}, 
and $h \to \tau^+ \tau^-$ \cite{atlastautau,Chatrchyan:2014nva}, 
while the CDF and $D\emptyset$ collaborations have analyzed the 
$\gamma\gamma$, $WW^*$ and $b\bar b$ decay modes of 
the Higgs boson \cite{Aaltonen:2013kxa,Junk:2014yea}. 
The experimental findings are usually presented in terms of the 
signal strength variable ($\mu$), which is defined as 
the ratio of the production cross section ($\sigma$) times the 
branching ratio ($\rm Br$) to a specific decay mode 
for a given new physics model normalized to the SM prediction. 
For example, when the Higgs boson is produced 
via gluon-gluon fusion process and it decays to a generic final 
state $X \bar X$ ($\gamma\gamma$, $WW^*$, $ZZ^*$, $b\bar b$ and 
$\tau^+ \tau^-$), then one can define the signal strength 
variable $\mu$, assuming narrow-width approximation, as:  
\bea \label{eq:R1}
\mu_{ggF}(X \bar X) = \frac{\Gamma(h\rightarrow gg )}{\Gamma(h_{SM}\rightarrow gg)}  
                      \times 
                      \frac {Br(h\rightarrow X \bar X)}{Br(h_{SM}\rightarrow X \bar X)}, 
\eea
where $h$ is a observed 125 GeV Higgs boson and $h_{SM}$ is the 
SM Higgs boson. Similarly, if the Higgs boson is produced via VBF fusion process and 
it decays to $X\bar X$, then one can define, 

\bea \label{eq:R2}
\mu_{VBF/VH}(X \bar X) = \frac{\Gamma(h\rightarrow WW )}{\Gamma(h_{SM}\rightarrow WW)}  
                      \times 
                      \frac {Br(h\rightarrow X \bar X)}{Br(h_{SM}\rightarrow X \bar X)}. 
\eea

From Eq.~\ref{eq:R1} and \ref{eq:R2} it is evident that the signal 
strengths are functions of the partial decay widths and the 
total decay width of the Higgs boson. The presence of new 
particles/interactions in the new physics models leads 
to modifications in the partial/total decay widths, and 
thereby changes in the value of signal strength variable. Thus, precise 
measurements of these signal strength variables are extremely 
crucial as a small but statistically significant deviation 
from the SM expectation will hint at possible signatures 
of the new physics. Note that, in the timeline of the 
Moriond 2014 \cite{moriond2014} workshop to the end of December 2014, significant 
changes have been observed in the measurement 
of various Higgs signal strengths. For example, ATLAS data 
for di-photon signal strength has 
changed from $1.57^{+0.33}_{-0.28}$ \cite{moriond2014} to 
1.17 $\pm$ 0.27 \cite{Aad:2014eha}, while 
the same from the CMS has changed 
from $0.77\pm 0.27$ \cite{moriond2014} to 
$1.14^{+0.26}_{-0.23}$ \cite{Khachatryan:2014ira}. Besides, both statistical 
and systematics uncertainties 
associated to some of these signal strengths have 
been reduced after the combination of (7+8) TeV 
LHC data. Near the end of last year, the 
CDF and $D\emptyset$ collaborations have also 
updated the constraints of 125 GeV Higgs boson 
couplings to fermions and vector bosons. In this work, 
we perform a global analysis considering all 
these most updated Higgs signal strengths as 
of the end of 2014. The available signal strength 
variables\footnote{For global fit analysis, we consider 
the signal strengths for different production modes as 
presented for individual decay channels. In other words, we 
do not consider the inclusive results for a given decay 
mode.}, for different production and decay modes 
of the Higgs boson at the LHC 
are summarized in Table~\ref{t1} to Table~\ref{t5}
while Tevatron Higgs data is shown in Table~\ref{t6}.

%-------------------- gamma gamma channel------------------
%--------------------------------------------------
\begin{table}[htb!]
\begin{center}
\begin{tabular}{|c | cc | ccc |}
\hline
Channel & \multicolumn{2}{c|}{Signal strength ($\mu$)}  & \multicolumn{3}{c|}{Production mode} \\
\cline{2-6}
        &  {\it ATLAS}               & {\it CMS}                & ggF & VBF & Vh \\
\hline
\hline
$\mu ({ggh})$ & $1.32\pm0.38$        &$1.12^{+0.37}_{-0.32}$    & 100\% & -     & - \\
$\mu ({VBF})$ & $0.8\pm 0.7$         & $1.58^{+0.77}_{-0.68}$   & -     & 100\% & - \\
$\mu ({Wh})$  & $1.0\pm1.6$          & $-0.16^{+1.16}_{-0.79}$  & -     & -     & 100\% \\
$\mu ({Zh})$  & $0.1^{+3.7}_{-0.1}$  & -                        & -     & -     & 100\% \\
\hline
\end{tabular}
  \caption{\small \label{t1}
Signal strengths of $h\rightarrow \gamma \gamma$ channel as recorded by the 
ATLAS \cite{Aad:2014eha} and CMS \cite{Khachatryan:2014ira} collaborations 
after 7+8 TeV run of LHC with 25 $\rm fb^{-1}$ of luminosity. The amount 
of contribution to a given channel from each production modes are shown in 
Column 4-6. The total $\chi^2$ for the $\gamma \gamma$ channel with respect to 
the SM is = 0.85 (ATLAS) + 1.868 (CMS) = 2.718.}
\end{center}
\end{table}
%--------------------------------------------------
%--------------------------------------------------
%-------------------------------ZZ-------------------
\begin{table}[htb!]
\begin{center}
\begin{tabular}{|c | cc | ccc |}
\hline
Channel & \multicolumn{2}{c|}{Signal strength ($\mu$)}  & \multicolumn{3}{c|}{Production mode} \\
\cline{2-6}
        &  {\it ATLAS} & {\it CMS}    & ggF & VBF & Vh \\
\hline
\hline
$\mu ({ggh+bbh+tth})$   & $1.66^{+0.51}_{-0.44}$ &$0.80^{+0.46}_{-0.36}$  & 100\%  & -          & - \\
$\mu ({VBF+Vh})$        & $0.26^{+1.64}_{-0.94}$ & $1.7^{+2.2}_{-2.1}$    & -      & 60\%       &40\%  \\
\hline
\end{tabular}
  \caption{\small \label{t2}
Signal strengths of $h\rightarrow ZZ^*$ channel as recorded by the 
ATLAS \cite{Aad:2014eva} and CMS \cite{Chatrchyan:2013mxa} collaborations 
after 7+8 TeV run of LHC with 25 $\rm fb^{-1}$ of luminosity. The amount 
of contribution to a given channel from each production modes are shown in 
Column 4-6. The total $\chi^2$ for the $ZZ^*$ channel with respect to 
the SM is = 2.493 (ATLAS) + 0.3 (CMS) = 2.793.}
\end{center}
\end{table}
%--------------------------------------------------

%----------------------------WW ----------------------
\begin{table}[htb!]
\begin{center}
\begin{tabular}{|c | cc | ccc |}
\hline
Channel & \multicolumn{2}{c|}{Signal strength ($\mu$)}  & \multicolumn{3}{c|}{Production mode} \\
\cline{2-6}
                                &  {\it ATLAS}          & {\it CMS} & ggF  & VBF   & Vh  \\
\hline
\hline
$\mu (ggF)$                     &$1.02^{+0.29}_{-0.26}$ & -        & 100\% & -     & -    \\
$\mu ({VBF})$                   &$1.27^{+0.53}_{-0.45}$ &  -       &     - & 100\% & -  \\
\hline
$\mu (0/1~jet)$                 &-      &$0.74^{+0.22}_{-0.20}$    &97\%  & 3\%  & -    \\
$\mu$ (VBF tag)                 &-      &$0.60^{+0.57}_{-0.46}$    &17\%  & 83\% & -    \\
$\mu$ (Vh tag $(2l2\nu 2j))$    &-      &$0.39^{+1.97}_{-1.87}$    & -    &  -    & 100\%  \\
$\mu$ (Wh tag$(3l3\nu))$        &-      &$0.56^{+1.27}_{-0.95}$    &-     & -    &  100\%       \\
\hline
\end{tabular}
  \caption{\small \label{t3}
Signal strengths of $h\rightarrow WW^*$ channel as recorded by the 
ATLAS \cite{ATLAS:2014aga} and CMS \cite{Chatrchyan:2013iaa} collaborations 
after 7+8 TeV run of LHC with 25 $\rm fb^{-1}$ of luminosity. The amount 
of contribution to a given channel from each production modes are shown in 
Column 4-6. The total $\chi^2$ for the $WW^*$ channel with respect to 
the SM is = 0.366 (ATLAS) + 2.104 (CMS) = 2.470.}
\end{center}
\end{table}
%--------------------------------------------------

%----------------------------bb ----------------------
\begin{table}[htb!]
\begin{center}
\begin{tabular}{|c | cc | ccc |}
\hline
Channel & \multicolumn{2}{c|}{Signal strength ($\mu$)}  & \multicolumn{3}{c|}{Production mode} \\
\cline{2-6}
                &{\it ATLAS}            & {\it CMS}     & ggF   & VBF   & Vh  \\
\hline
\hline
$\mu$(Vh tag)   &$0.51^{+0.40}_{-0.37}$ & $1.0 \pm 0.5$ & -     & -     & 100\%    \\
\hline
\end{tabular}
  \caption{\small \label{t4}
Signal strengths of $h\rightarrow b \bar b$ channel as recorded by the 
ATLAS \cite{Aad:2014xzb} and CMS \cite{Chatrchyan:2013zna} collaborations 
after 7+8 TeV run of LHC with 25 $\rm fb^{-1}$ of luminosity. The amount 
of contribution to a given channel from each production modes are shown in 
Column 4-6. The total $\chi^2$ for the $b\bar b$ channel with respect to 
the SM is = 1.50 (ATLAS) + 0.0 (CMS) = 1.5.}
\end{center}
\end{table}

%--------------------------------------------------
%----------------------------tau tau ----------------------
\begin{table}[htb!]
\begin{center}
\begin{tabular}{|c | cc | ccc |}
\hline
Channel & \multicolumn{2}{c|}{Signal strength ($\mu$)}  & \multicolumn{3}{c|}{Production mode} \\
\cline{2-6}
        &  {\it ATLAS} & {\it CMS}    & ggF & VBF & Vh \\
\hline
\hline
$\mu ({ggF})$ 	&$1.93^{+1.45}_{-1.15}$ & -     & 100 \%  &  - & - \\
$\mu ({VBF+Vh})$&$1.24^{+0.58}_{-0.54}$ &  -	  & - 	& 60\% & 40\%  \\
\hline
$\mu$ (0-jet)	&-	&$0.34 \pm 1.09$    &96.9\%  & 1.0\% & 2.1    \\
$\mu$ (1-jet)	&-	&$1.07 \pm 0.46$    &75.7\% & 14\% & 10.3    \\
$\mu$ (VBF tag)	&-	&$0.94 \pm  0.41$    & 19.6 	& 80.4    & -\\
$\mu$ (Vh tag)	&-	&$-0.33 \pm 1.02$    &-  & - &  100\% 	\\
\hline
\end{tabular}
  \caption{\small \label{t5}
Signal strengths of $h\rightarrow \tau^{+} \tau^{-}$ channel as recorded by the 
ATLAS \cite{atlastautau} and CMS \cite{Chatrchyan:2014nva} collaborations 
after 7+8 TeV run of LHC with 25 $\rm fb^{-1}$ of luminosity. The amount 
of contribution to a given channel from each production modes are shown in 
Column 4-6. The total $\chi^2$ for the $\tau^{+}\tau^{-}$ 
channel with respect to the SM is = 0.857 (ATLAS) + 2.11 (CMS) = 2.967.}
\end{center}
\end{table}
%--------------------------------------------------

%----------------------------Tevatron ----------------------
\begin{table}[htb!]
\begin{center}
\begin{tabular}{|c | c | ccc |}
\hline
Channel & {Signal strength ($\mu$)}  & \multicolumn{3}{c|}{Production mode} \\
\cline{2-5}
        & {\it Tevatron}  & ggF & VBF & Vh  \\
\hline
\hline
$\mu(H \to \gamma \gamma)$   &  $6.14^{+3.25}_{-3.19}$ & 78\% & 5\% & 17\%    \\
$\mu(H \to WW^*)$   &  $0.85^{+0.88}_{-0.81}$  & 78\% & 5\% & 17\%    \\
$\mu(H \to b\bar b)$   & $1.59^{+0.69}_{-0.72}$ & - & - & 100\%    \\
\hline
\end{tabular}
  \caption{\small \label{t6}
Signal strengths of $h\rightarrow \gamma\gamma, WW^* , {\rm and} 
~b \bar b$ channel as recorded by the 
CDF and $D\emptyset$ collaborations at the Tevatron with 10 $\rm fb^{-1}$ of 
luminosity at $\sqrt s$ = 1.96 TeV \cite{Aaltonen:2013kxa,Junk:2014yea}. 
The amount of contribution to a given channel from each production 
modes are shown in Column 4-6. The total $\chi^2$ for the above three modes 
with respect to the SM is = 3.296.}
\end{center}
\end{table}

%% Ref: Aurelio Juste, “Standard Modek Higgs boson searches at the Tevatron ”, talk at HCP2012, 
%% 15 Nov 2012, Kyoto, Japan,
%% http://kds.kek.jp/conferenceDisplay.py?confId=9237.
%% Yuji Enari, “H → bB from Tevatron”, talk at HCP2012, 14 Nov 2012, Kyoto, Japan,
%% http://kds.kek.jp/conferenceDisplay.py?confId=10808.

%--------------------------------------------------

Let us now discuss the details of the MSSM parameter space scan. 
We do not consider any specific SUSY breaking scenario, rather 
we focus on the generic phenomenological MSSM (pMSSM) model 
with 19 free parameters and perform a random scan for 
approximately 100 million points. Those parameters that 
are relevant to the MSSM Higgs sector, namely pseudo-scalar 
mass parameter $M_A$, the ratio of the vacuum 
expectation values of two Higgs doublets $\tan\beta$, higgsino mass parameter 
$\mu$, the third generation squark trilinear couplings 
$A_t$ and $A_b$ (trilinear couplings of sleptons and first 
two generations squarks are set to zero), third 
generation squark soft mass parameters $M_{Q3}$, $M_{U3}$ and $M_{D3}$, are 
scanned in the following ranges:
\begin{eqnarray}
\label{eq:S}
1 < \tan\beta < 50,&\ \  100~ {\rm GeV} < M_{A} < 600~ {\rm GeV}, \nonumber \\
-8000~ {\rm GeV} < A_{\rm t}, A_{\rm b} < 8000~ {\rm GeV}, & \ \ 100~ {\rm GeV} <~\mu ~~ < 8000~ {\rm GeV}, \nonumber \\
100~ {\rm GeV} < M_{\rm Q3},~M_{\rm U3} < 8000~ {\rm GeV}, & 100~ {\rm GeV} < ~M_{\rm D3} < 8000~ {\rm GeV}, 
\label{parameterRanges}
\end{eqnarray}
while we fix the following parameters since they have little impact 
on our analysis,  
\begin{eqnarray}
M_1 = 100~{\rm GeV}, & M_2 = 2000~{\rm GeV}, &  M_3 = 3000~{\rm GeV},  \nonumber \\
M_{\rm L_{1,2,3}} = M_{\rm E_{1,2,3}} = 3000~{\rm GeV}, & 
M_{\rm Q_{1,2}} =  3000~{\rm GeV}, & M_{\rm U_{1,2}} = M_{\rm D_{1,2}} = 3000~{\rm GeV},    
\label{parameterRanges}
\end{eqnarray}
where $M_{1,2,3}$ are the gaugino mass parameters, $M_{\rm L_{i}}$ and 
$M_{\rm E_{i}}$ ~($i = 1,2,3$) are the left and 
right handed slepton soft SUSY breaking mass parameters, and 
$M_{\rm Q_{i}}$, $M_{\rm U_{i}}$, $M_{\rm D_{i}}$ ~($i = 1,2$) 
are the first two generation squark soft SUSY breaking mass parameters.   

We scan the third generation trilinear couplings 
($A_t$ and $A_b$) and soft masses ($M_{\rm Q3},~M_{\rm U3}, M_{\rm D3}$) 
over a wide range in order to obtain the lightest 
MSSM Higgs boson mass in the range of 
125 $\pm$ 3 GeV assuming 3 GeV uncertainty 
in Higgs mass calculation \cite{higgsuncertainty3GeV}. Since 
we are interested in the possibility of having 
light additional Higgses, we restrict $M_A$ up to 600 GeV. 
From the above choices of the model parameters, it is evident 
that we do not consider the possibility of the decay of $h$ to 
MSSM particles. We use SUSPECT (version 2.43) \cite{Djouadi:2002ze} to 
scan the MSSM parameter space and SuperIso (version 3.4) \cite{superiso} 
to calculate the flavour physics observables, while the 
branching ratios of the lightest Higgs boson are evaluated using 
HDECAY (version 6.41) \cite{Djouadi:1997yw}.

Now, to combine the available information on different signal
strength variables from the ATLAS, CMS and Tevatron, and to compare 
with the MSSM expectations, we compute $\chi^2$ for all the scanned 
parameter space points, defined as below: 
\begin{equation}
  \chi^2=\sum_i {(\overline\mu_i-\mu_i)^2\over \Delta \mu_i^2}\,,
\label{chisqdef}
\end{equation}
where $\mu_i$ is the experimentally observed signal strength 
for a particular production/decay 
mode $i$, and $\overline\mu_i$ is the value predicted for the same 
channel for a chosen MSSM parameter space point with 
$\Delta \mu_i$ being the measure of the experimental error 
associated to that channel. The sum over $i$ takes into account 
of all the experimentally measured production and 
decay modes of the Higgs boson.

It is to be noted that, different production processes can, in 
principle, contribute to a particular experimental search channel, 
thereby while calculating the signal strengths for a parameter 
space point, contributions coming from different production 
processes need to be considered. Following the 
procedure of Ref.~\cite{Belanger:2012gc}, we implement this 
modification in our analysis as follows: 
\begin{equation}
  \overline\mu_i=\sum T^j_i \widehat \mu_j,
  \label{murel}
\end{equation}
where $T^j_i$ denotes the amount of contribution that 
can originate from the production mode $j$ to the category/channel 
$i$ with $\widehat \mu_i$ being the signal strength corresponding to 
the MSSM parameter space point. For example, the 
category $\mu(\rm {VBF tag})$, as introduced in Table~\ref{t3}, 
receives 17\% and 83\% contributions from the ggF and VBF processes 
respectively \cite{lhchiggsxsec}. So, we calculate $\widehat \mu_{ggF}$ and 
$\widehat \mu_{VBF}$, and then scale them with 0.17 and 0.83 
(the $T^j_i$s here) respectively to obtain the proper signal 
strength ($\overline\mu$) corresponding to ${\rm VBF tag}$ 
category. To obtain the contributions coming 
from different production processes to a given decay mode/category, 
we use LHC Higgs cross section working group report \cite{lhchiggsxsec}.

We consider altogether 28 data points (i.e., experimental inputs, 
see Table~\ref{t1} - \ref{t6}) combining 
the CMS and ATLAS and Tevatron Higgs data. We calculate $\chisq$ for 
all the scanned parameter space points and find the minimum of 
$\chisq$. We call this minimum as the {\it approximate} 
minima ($\chi^2_{\rm approx}$). In order to obtain the {\it true} 
$\chisq$ minimum ($\chi^2_{\rm min}$), we vary the parameters around 
their {\it approximated} values i.e., values corresponding 
to the approximate $\chisq$ minimum. We present the parameter space 
that is available after the global analysis considering 
the 1$\sigma$ and 2$\sigma$ intervals with 
$\chisq=\chi^2_{\rm min}+2.3$ and $\chisq=\chi^2_{\rm min}+6.18$ 
respectively in $M_{A} - \tan\beta$ plane \cite{PDG}. 
The ``best fit" value corresponds to $\rm {M_{A} \sim 584 ~GeV }$ and 
$\rm {\tan\beta} \sim ~36$. 
To determine how a set of experimental data is well represented by any given 
model, one usually calculates the chi-square per 
degrees of freedom ($\rm d.o.f$) i.e., $\chisq/{\rm d.o.f}$.   
The minimum value of $\chisq$ obtained from the analysis for SM 
is 15.744 with $\chisq/{\rm d.o.f}$ = $\chisq/28$  = 0.562, while for MSSM we 
obtain $\chisq_{\rm min}$= 15.013 with $\chisq/{\rm d.o.f}$ = $\chisq/20$ = 0.75. 
%One can also express these results in terms of $p$-values\footnote{The 
%$p$-value, the measure of the goodness of a 
%fit, is given by \cite{PDG} 
%\bea
% p &=& \int_{\chi^2}^{\infty} f(x;n) dx \nonumber \\
%   &=& \int_{\chi^2}^{\infty} \frac{x^{n/2-1} e^{-x/2} }{ 2^{n/2} \Gamma(n/2) } dx \nonumber 
%\eea
%where $n$ is the degrees of freedom.
%}
%for the SM we find $p_{\rm SM}$ = ..., while for MSSM 
%we get $p_{\rm pMSSM}$ = .... 

\subsection{Available MSSM parameter space}
\label{sec21}

%=============================================

Before we proceed to discuss our findings, let us review our 
methodology once more. We first perform a $\chisq$ analysis 
using a random scan of the parameters relevant to the MSSM Higgs 
sector, then we find the true $\chisq_{\rm min}$ to get 
the ``best-fit" values of those parameters. The points 
with $\chisq$ within $2\sigma$ of the true $\chisq_{\rm min}$ 
are only considered for further analysis. We impose 
the two most stringent rare b-decay constraints, namely 
$\rm {Br (b \to s \gamma)}$ and $\rm {Br (B_{s} \to \mu^+ \mu^-)}$, 
and allow $2\sigma$ deviation\footnote{The current measurements of these 
two b-observables are ${\rm Br}(B_s \to X_{s}\gamma) = 3.43 \pm 0.22 \pm 0.21({\rm theo.})$ 
and ${\rm Br}(B_s \to \mu^+ \mu^-) = 3.1 \pm 0.7 \pm 0.31 ({\rm theo.})$ \cite{Amhis:2014hma}. 
We follow Ref.~\cite{Kowalska:2015zja} for the conservative estimates of the theoretical 
uncertainties associated to these two flavor observables.} \cite{Amhis:2014hma}, 
\bea
 \rm 2.82\times 10 ^{-4} < Br(B_s \to X_{s}\gamma) < 4.04\times 10^{-4} \nonumber \\ 
 \rm 1.57\times 10 ^{-9} < Br(B_s \to \mu^+ \mu^-) < 4.63\times10^{-9}.  
\eea

\begin{figure}[!htb]
 \begin{center}
{\includegraphics[angle =0, width=0.6\textwidth]{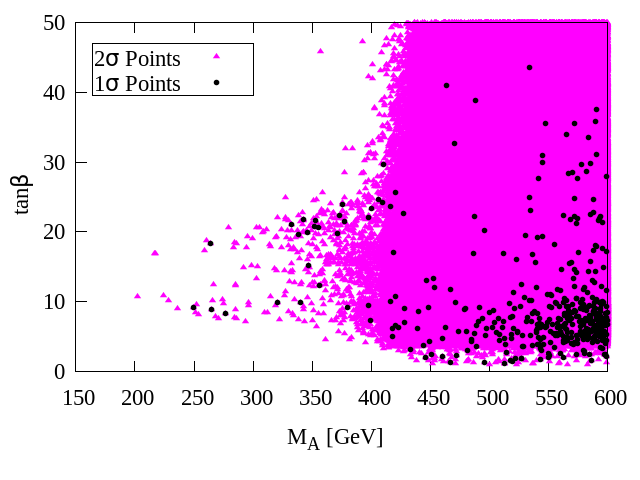} }
\caption{ {\it Parameter space allowed in $M_A$ - tan$\beta$ plane from our global 
fit analysis and also satisfying the flavour physics constraints on Br(b $\rightarrow s \gamma$) 
and Br($B_s \rightarrow \mu^+ \mu^-$). Magenta (black) coloured triangle (circle) shaped 
points represent 2$\sigma$ (1$\sigma$) allowed parameter space from global fits of 
125 GeV Higgs data after Run-I of LHC. In rest of our analysis, while presenting 
the direct search constraints and the future limits on the heavy Higgs 
masses and BRs, we would consider these 2$\sigma$ points which satisfies 
our global fit analysis and the updated flavour data.}}
\label{fig:ma_tb1}
 \end{center}
\end{figure}

In Fig.\ref{fig:ma_tb1}, we show the parameter space 
in the $\rm {M_{A} - \tan\beta}$ plane obtained 
from the global fit analysis and also satisfying the 
flavour physics constraints. Magenta (black) coloured 
triangle (circle) shaped points represent 2$\sigma$ (1$\sigma$) allowed 
parameter space. The region with $\rm {M_{A} \le 350 ~GeV}$ 
and $\rm {\tan\beta} \ge 25$ are excluded by the stringent 
$\rm {Br (B_{s} \to \mu^+ \mu^-)}$ constraint, which is expected 
to dominate for regions with large $\rm {\tan\beta}$ and 
relatively smaller $\rm {M_A}$. However, for significantly 
larger $\rm {M_A}$ i.e., $\rm {M_{A} \ge 425 ~GeV }$, the 
effect of this constraint is negligible. Besides, most of the 
points in the region with $\rm {M_{A} \le 350 ~GeV }$ 
with $\rm {\tan\beta} \le ~8$ 
are excluded by the $\rm {Br (b \to s \gamma)}$ constraint. 
%One 
%can understand this effect as follows: smaller values of ${\rm M_A}$ 
%implies relatively light charged Higgses which can induce 
%substantial enhancement in the $b \to s \gamma$ branching ratio, 
%thus regions with light ($<$ 300 GeV) charged bosons 
%are thereby excluded. 

%--------------------------------------------------
%--------------------------------------------------

%--------------------------------------------------
\begin{figure}[!tb]
%\begin{center}
\includegraphics[angle =0, width=0.32\textwidth]{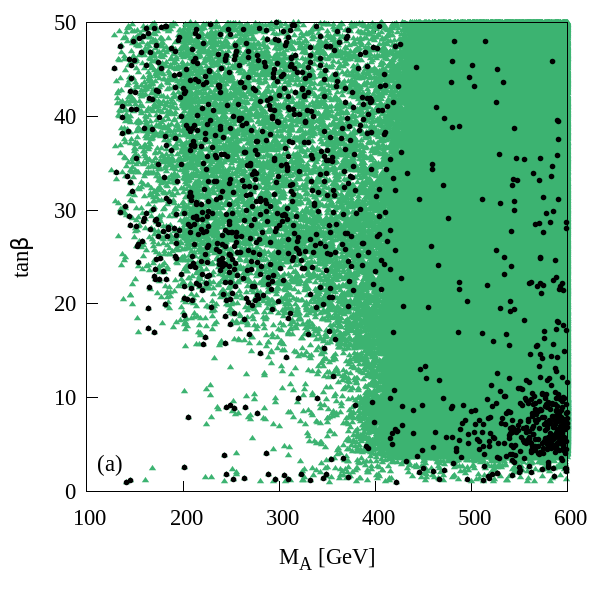}
\includegraphics[angle =0, width=0.32\textwidth]{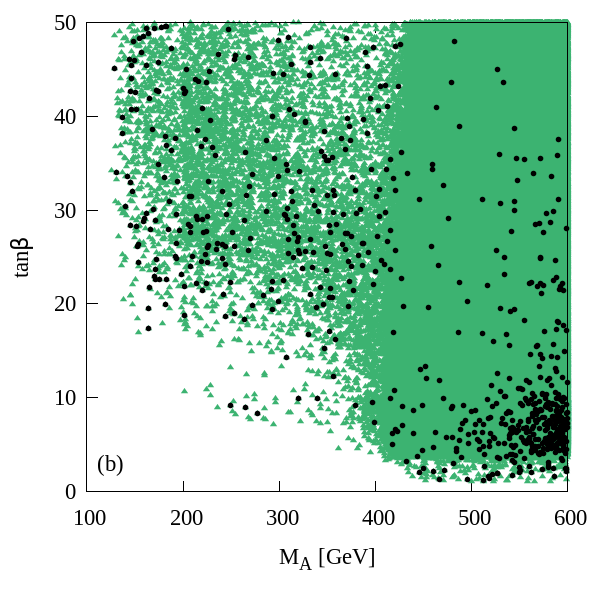} 
\includegraphics[angle =0, width=0.32\textwidth]{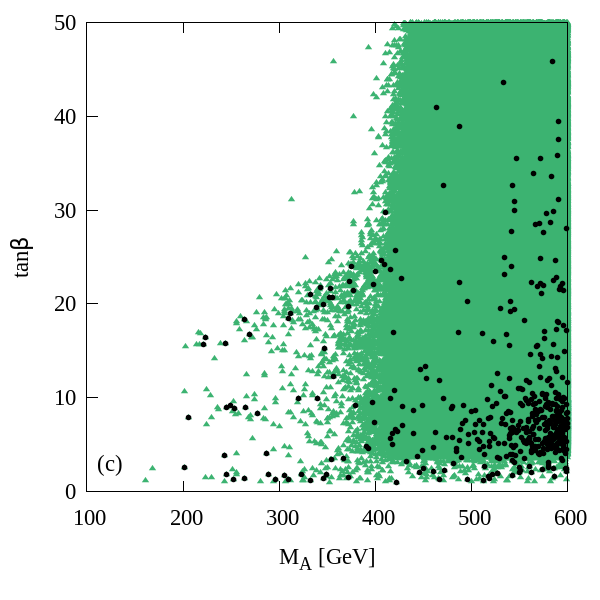}
\caption{ {\it Scatter plots in the (a) $M_{A} - \tan\beta$ plane without any 
flavor constraint (left panel), (b) $M_{A} - \tan\beta$ plane only after 
imposing Br(b $\rightarrow s \gamma$) constraint (middle panel) and 
(c) $M_{A} -\tan\beta$ plane after imposing only 
Br($B_s \rightarrow \mu^+ \mu^-$) constraint (right panel). In 
Fig.~\ref{fig:ma_tb1}, we show the same correlation after the 
imposition of both Br(b $\rightarrow s \gamma$) and 
Br($B_s \rightarrow \mu^+ \mu^-$) constraints. }}
\label{fig:ma_tb2}
%\end{center}
\end{figure}
%--------------------------------------------------

In order to display the interplay of two flavour physics 
constraints, in Fig.~\ref{fig:ma_tb2}($a$) we first show the parameter
space obtained after the global analysis with Higgs mass constraint 
($122 ~{\rm GeV} \le M_{h} \le 128 ~{\rm GeV}$) only, without 
imposing any flavor physics constraint. 
In the middle panel ($b$), we then show the 
same distribution but only after imposing the 
$\rm {Br (b \to s \gamma)}$ constraint.  In right-most figure 
(Fig.~\ref{fig:ma_tb2}$c$) we plot the same correlation with  
imposing $\rm {Br (B_{s} \to \mu^+ \mu^-)}$ constraint only. 
The effect of the $\rm {Br (B_{s} \to \mu^+ \mu^-)}$ 
constraint in the low $M_A$ and large $\tan\beta$ region, 
and the impact of $\rm {Br (b \to s \gamma)}$ in low $M_A$ 
and low $\tan\beta$ is now clearly visible from these plots. 
Once we impose both the Higgs mass and flavour physics
constraints, the available parameter space has already been shown 
in Fig.~\ref{fig:ma_tb1}. Note that, in rest of our 
analysis, we name these 2$\sigma$ allowed 
points collectively as the ``scanned data set" and present 
the direct search constraints and the future limits on 
the heavy Higgs masses and couplings using this data set.

%--------------------------------------------------
\begin{figure}[!tb]
\begin{center}
%        \subfigure[]{ 
{\includegraphics[angle =0, width=0.48\textwidth]{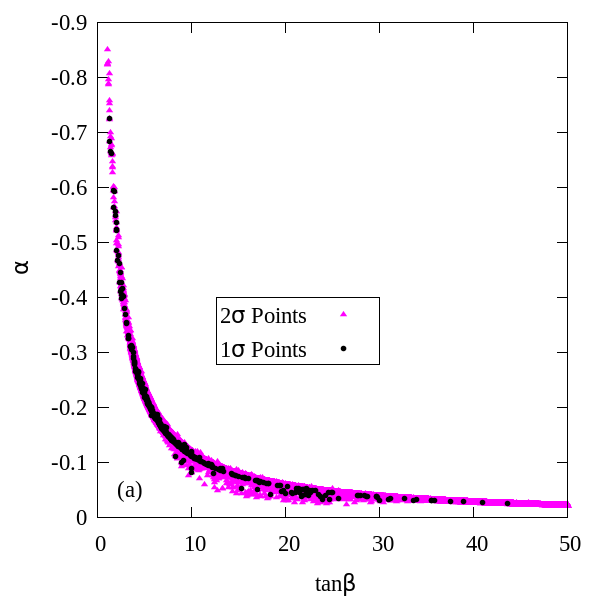} }
%} 
%        \subfigure[]{
{\includegraphics[angle =0, width=0.48\textwidth]{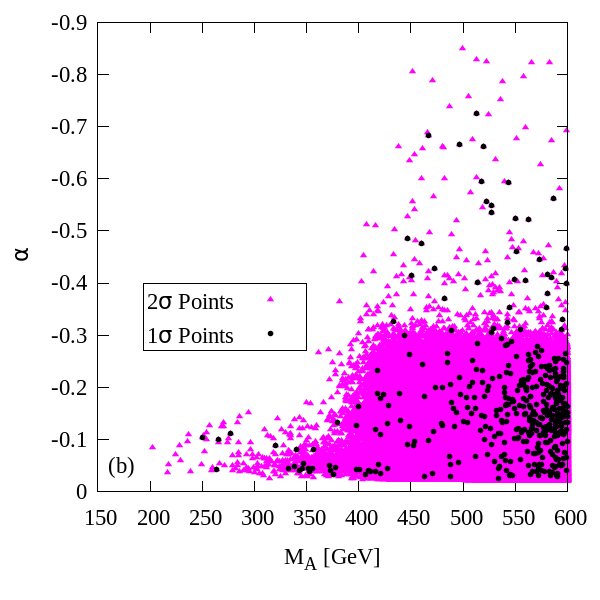} }
%}
%        \subfigure[]{
%\includegraphics[angle =270, width=0.31\textwidth]{plots/ma_tb_after_bsmumu.pdf}}
%}
\caption{ {\it Left panel (a) shows the allowed parameter space 
in $\tan\beta - \alpha$ plane after the global fit analysis and 
also satisfying the flavour physics constraints. 
% on Br(b $\rightarrow s \gamma$) and Br($B_s \rightarrow \mu^+ \mu^-$). 
Colour conventions are same as Fig.~\ref{fig:ma_tb1}. 
In the right panel (b), we display the same allowed parameter 
space but now is presented in the $M_{A} - \alpha$ plane. }}
\label{fig:alp_tb}
\end{center}
\end{figure}
%--------------------------------------------------

\begin{figure}[!htb]
\begin{center}
\includegraphics[angle =0, width=0.6\textwidth]{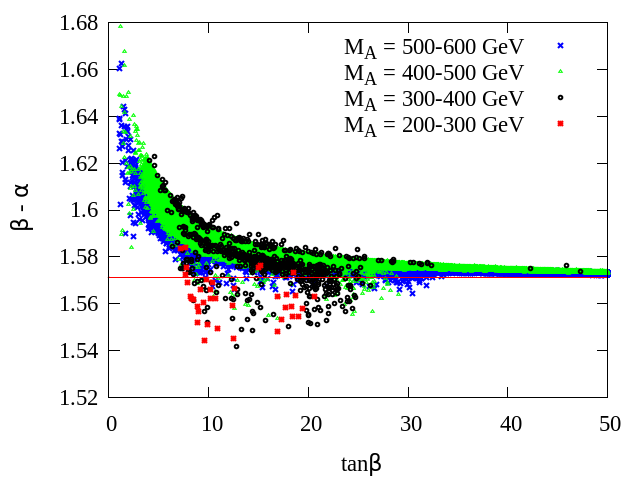}
%{\includegraphics[angle =270, width=0.48\textwidth]{plots/tb_alpha.pdf} }
%{\includegraphics[angle =270, width=0.48\textwidth]{plots/ma_alpha.pdf} }
%{\includegraphics[angle =270, width=0.48\textwidth]{tb_beta-alpha.pdf} }
\caption{ {\it The allowed parameter space in the 
$(\beta - \alpha) - \tan\beta$ plane is presented with different
choices of $M_A$. The entire $M_{A}$ region is divided 
in four parts, namely $200 < M_{A} < 300$~{\rm GeV} (red/square),
$300 < M_{A} < 400$~{\rm GeV} (black/circle),
$400 < M_{A} < 500$~{\rm GeV} (green/triangle), and
$500 < M_{A} < 600$~{\rm GeV} (blue/cross). The red horizontal 
line at $(\beta - \alpha) = 1.571$ indicates the value 
of $(\alpha - \beta)$ at the alignment limit. We see that regions 
with light $M_A$ ($\le$ 400 GeV) satisfying the alignment 
limit are perfectly allowed by the current data.}}
\label{fig:alp_tb2}
\end{center}
\end{figure}

In Fig.~\ref{fig:alp_tb}, we show the scatter plots 
in the $(a)$ $\alpha - \tan\beta$ and $(b)$ 
$\alpha - M_{A}$ planes, where $\alpha$ is 
the Higgs mixing angle. But, before we proceed, let's 
first briefly discuss the idea of the ``alignment limit" 
(for details see 
Refs.\cite{Gunion:2002zf,Delgado:2013zfa,Craig:2013hca,Carena:2013ooa}). 
In the MSSM, the couplings (at the tree level) of 
the CP-even Higgs bosons ($h,H$) to 
SM gauge 
bosons are \cite{Djouadi:2005gj}, 
\bea
g_{hVV} = \sin(\beta - \alpha)~g_{V} \nonumber \\ 
g_{HVV} = \cos(\beta - \alpha)~g_{V}, 
\label{eq:decop1}
\eea   
while, couplings to the SM fermions 
are \cite{Djouadi:2005gj},
\bea
g_{hdd} = -\sin\alpha/\cos\beta g_{f} = (\sin(\beta - \alpha) - \tan\beta \cos(\beta - \alpha))~g_{f} \nonumber \\ 
g_{huu} = -\cos\alpha/\sin\beta g_{f} = (\sin(\beta - \alpha) + \cot\beta \cos(\beta - \alpha))~g_{f} \nonumber \\ 
g_{Hdd} = -\cos\alpha/\cos\beta g_{f} = (\cos(\beta - \alpha) + \tan\beta \sin(\beta - \alpha))~g_{f} \nonumber \\ 
g_{Huu} = -\sin\alpha/\sin\beta g_{f} = (\cos(\beta - \alpha) - \cot\beta \sin(\beta - \alpha))~g_{f},
\label{eq:decop2}
\eea   
where $g_V$ and $g_f$ are the corresponding SM couplings with 
$g_{V} = 2{i} M_{V}^2/v$ and $g_{f} = {i} M_{f}/v$ for a generic 
gauge boson $V$ ($V \equiv W,Z$) and a 
fermion $f$ with $v$ = 246 GeV. Now, the alignment limit is the limiting case 
when the lightest CP-even Higgs boson mimic the properties 
of the SM Higgs and the SM gauge bosons couple to the light 
CP-even Higgs boson only ( i.e., $g_{hVV} \sim 1$ and 
$g_{HVV} \sim 0$). This limit can be easily achieved with 
the variation of the two quantities $\alpha$ and $\beta$. 
For example, let's assume a special case: 
$(\beta - \alpha) \sim \pi/2$ for some specific 
choices of $\alpha$ and $\beta$, then the couplings of the 
lightest CP even Higgs boson to the SM gauge bosons are SM-like, while heavier 
CP even Higgs boson couplings become highly suppressed. Note 
that, even though the heavier CP even Higgs has highly 
suppressed coupling to the SM gauge bosons, however it can have 
non-zero couplings to the SM fermions depending on the choice 
of $\tan\beta$ (see Eq.~\ref{eq:decop2}). Now, let's discuss how 
we achieve the alignment limit from our parameter space scan. 
To do so, let's first divide the region of interest into three parts, 
namely $\tan\beta<5$, $5<\tan\beta<20$, and $\tan\beta>20$. 
The region with 
$\tan\beta<5$ corresponds to $\alpha >$ - 0.2 radian. 
Now, $\tan\beta = $ 5 implies $\beta =$ 1.373 radian, 
which means $(\beta - \alpha) = $ 1.573 radian or, 
around 90 degree with $\alpha =$ - 0.2 radian, so here 
we are near the alignment limit ($\beta - \alpha = \pi/2$) 
\cite{Gunion:2002zf}. The region with $\tan\beta>20$ corresponds 
to very small $\alpha$ ($<$ - 0.05 radian), 
and thereby  $(\beta - \alpha) \sim {\pi/2} $ 
i.e., we again achieve the alignment limit.
The intermediate regime with 
$5<\tan\beta<20$ also satisfies the criteria 
of alignment. However, we see that in 
this alignment limit, $M_A$ can be as 
light as 300 - 400 GeV with relatively 
large $\alpha$ (see Fig.~\ref{fig:alp_tb}$(b)$) satisfying 
current data. This scenario can be 
thought of as the ``alignment without decoupling" 
scenario, as discussed in 
Ref.~\cite{Delgado:2013zfa,Craig:2013hca,Carena:2013ooa}. In fact, 
in order to make a more quantitative statement, In 
Fig.~\ref{fig:alp_tb2} 
we present the allowed parameter space in 
the $(\beta - \alpha) - \tan\beta$ plane with different 
choices of $M_A$. We divide the entire $M_{A}$ region 
into four parts: $200 < M_{A} < 300$~{\rm GeV} (red/square), 
$300 < M_{A} < 400$~{\rm GeV} (black/circle), 
$400 < M_{A} < 500$~{\rm GeV} (green/triangle), and 
$500 < M_{A} < 600$~{\rm GeV} (blue/cross), while the red 
horizontal line at $(\beta - \alpha) = 1.571$ indicates the exact 
value at the alignment limit. From Fig.~\ref{fig:alp_tb} and 
Fig.~\ref{fig:alp_tb2} it is clear that 
regions with light $M_A$ ($\le$ 400 GeV) satisfying 
the alignment limit is perfectly allowed by the current data, one 
is thus not always forced to be in the decoupling limit to comply 
with LHC data.

%--------------------------------------------------
%--------------------------------------------------

%--------------------------------------------------
\begin{figure}[!htb]
 \begin{center}
 {\includegraphics[angle =0, width=0.32\textwidth]{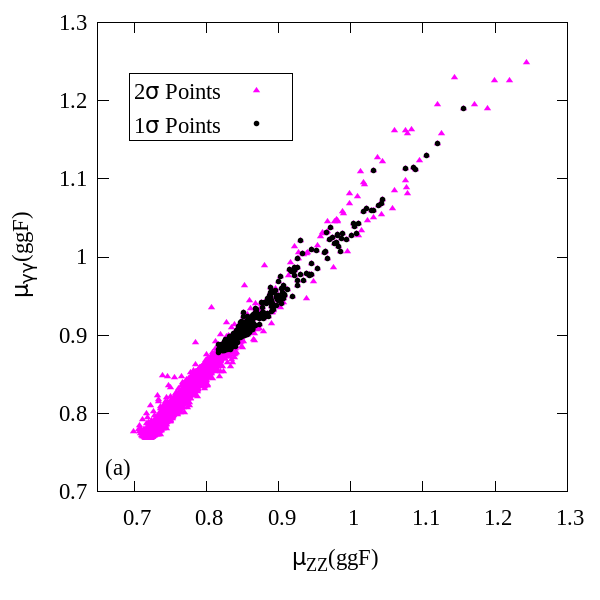} }
 {\includegraphics[angle =0, width=0.32\textwidth]{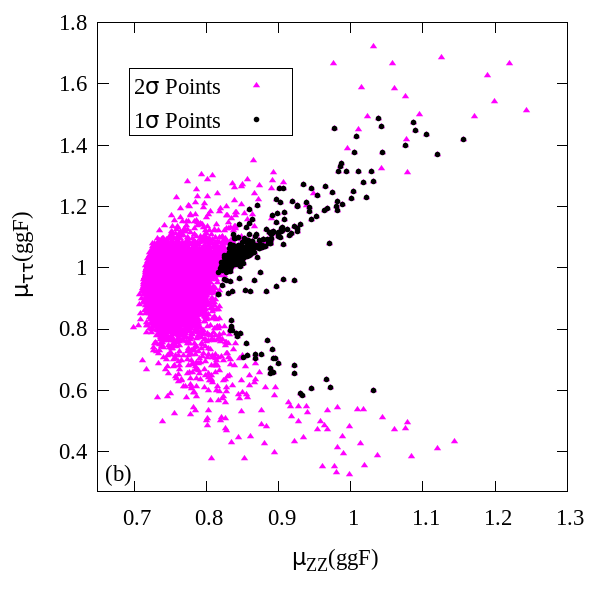} }
  {\includegraphics[angle =0, width=0.32\textwidth]{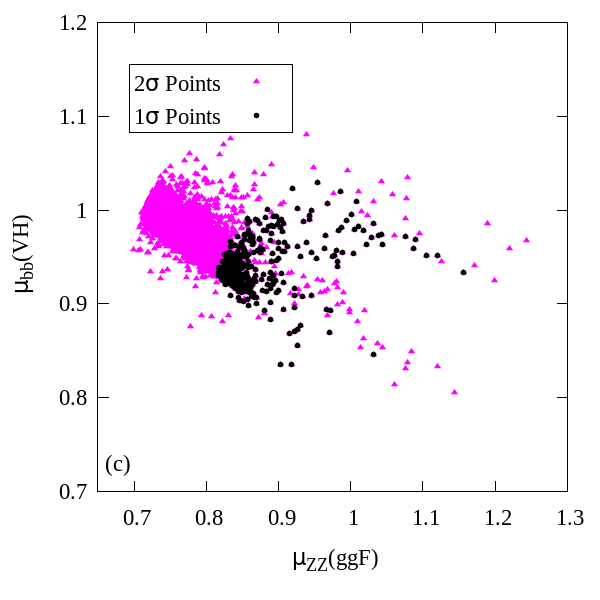} }

\caption{{\small {\it Signal strength correlations in  
(a) $\mu_{ZZ} - \mu_{\gamma\gamma}$, 
(b)$\mu_{ZZ} - \mu_{\tau^{+}\tau^{-}}$ and 
(c) $\mu_{ZZ} - \mu_{b\bar b}$ planes. 
Magenta (black) coloured triangle (circle) shaped 
points represent 2$\sigma$ (1$\sigma$) allowed points of our  ``scanned data set". 
}}}
\label{fig:mu}
 \end{center}
\end{figure}
%--------------------------------------------------

\begin{figure}[!htb]
 \begin{center}
 {\includegraphics[angle =0, width=0.32\textwidth]{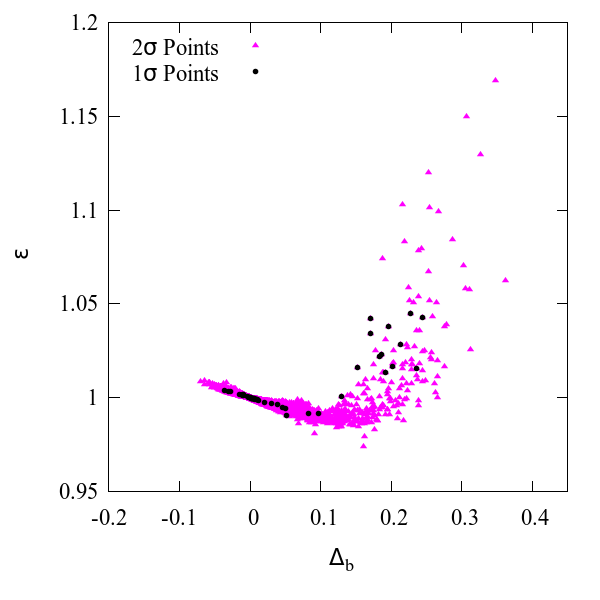} }
  {\includegraphics[angle =0, width=0.32\textwidth]{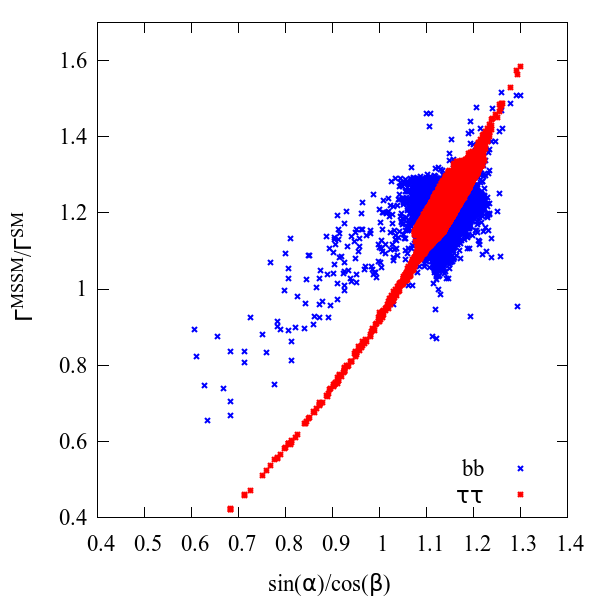} }
 {\includegraphics[angle =0, width=0.32\textwidth]{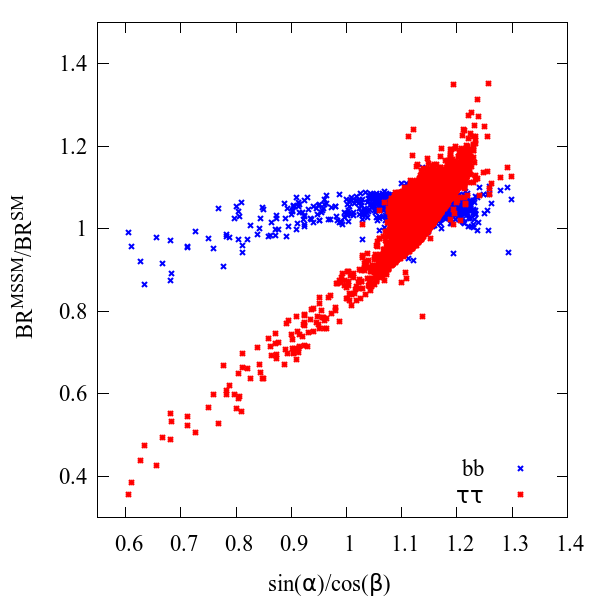} }
\caption{{\small Left panel shows the distribution of the quantity 
$\epsilon = \biggl({1\over 1+\Delta_b}\biggr) \times \biggl(1-{\Delta_b\over \tan\beta \tan \alpha}\biggr)$
with the variation of $\Delta_b$. The variation of the ratios 
of the partial decay widths and branching ratios of the Higgs boson 
to the SM values for the $b\bar b$ and $\tau^{+}\tau^{-}$ 
modes with the tree level Higgs Yukawa couplings 
$\frac{\sin\alpha}{\cos\beta}$ are shown in the middle and right 
panel respectively. For more details, see the text. }}
\label{fig:delb}
 \end{center}
\end{figure}

Before we end this section, we would like to discuss the 
correlations of various Higgs signal strength variables. 
In Fig.~\ref{fig:mu}, we show the scatter plots in 
$\mu_{ZZ}- \mu_{\gamma\gamma} $ (left), 
$\mu_{ZZ} - \mu_{\tau^{+}\tau^{-}}$ (middle) and 
$\mu_{ZZ} - \mu_{bb} $ (right) planes. We find that 
the partial widths $\Gamma(h \to ZZ)$ and 
$\Gamma(h \to WW)$ remain almost unaltered with 
respect to the SM 
values, however  
$\Gamma(h \to b\bar b)$ and total decay width of Higgs 
($\Gamma_{tot}$) have increased while $\Gamma(h \to gg)$ 
has decreased for most of the scanned data points. 
Now, we know that when the total decay width 
increases, the branching ratios of various 
sub-leading decay modes like 
$\gamma\gamma$, $gg$, $ZZ$, $WW$ decreases. Hence, the 
suppression in the $\mu_{ZZ}$/$\mu_{\gamma\gamma}$ can be thought 
of as the interplay of two effects: the increase in total decay 
width (or, decrease in 
$h \to ZZ$ or $h \to \gamma\gamma$ branching ratios) 
and decrease in $\Gamma(h \to gg)$ with respect to the SM. Now, 
in order to understand the $bb$ and $\tau\tau$ correlations in 
Fig.~\ref{fig:mu}(b) and Fig.~\ref{fig:mu}(c), we need to discuss 
the couplings of the lightest CP-even Higgs boson with 
the $b$ quark and $\tau$ lepton. At the tree level, 
the Yukawa couplings of $b$ and $\tau$ are proportional 
to $\frac{\sin\alpha}{\cos\beta}$ 
(see Eq.~\ref{eq:decop2}). However, loop 
corrections (in powers of $\alpha_s\tan\beta$) 
involving heavier supersymmetric particles can 
significantly modify the $b$ quark 
mass and it's Yukawa coupling from its tree level 
predictions. In an effective Lagrangian approach, these 
effects are parametrized by the quantity $\Delta_b$ and 
one can write \cite{Carena:1999py,Hall:1993gn,Guasch:2003cv,Dawson:2011pe},
\begin{equation}
L_{hb\bar b}=-{m_b\over v_{SM}}\biggl({1\over 1+\Delta_b}\biggr)
\biggl(-{\sin \alpha \over \cos\beta}\biggr)\biggl(1-{\Delta_b\over \tan\beta
\tan \alpha}\biggr) b{\overline b} h 
\label{eq:leff}
\end{equation}
where $v_{SM}=(\sqrt{2}G_F)^{-1/2}$. The bottom quark mass can now 
be written as,
\begin{equation}
m_b \to {y_{b} v_1\over \sqrt{2}} (1+\Delta_b),
\end{equation}
where $v_1=v_{SM}\cos\beta$ and $y_b$ the bottom Yukawa coupling. 
In order to understand various terms 
of Eq.~\ref{eq:leff}, in Fig.~\ref{fig:delb}(a), we 
show the distribution of the quantity,
\beq 
\epsilon = \biggl({1\over 1+\Delta_b}\biggr) \times \biggl(1-{\Delta_b\over \tan\beta \tan \alpha}\biggr) 
\eeq
with respect to $\Delta_b$ for all the 
2$\sigma$ (magenta triangle) and 1$\sigma$ (black circles) allowed 
points. It is evident from the figure that for significant number 
of points, $\Delta_b$ is mostly positive and varies within 
10-15\%. However, the effect of this variation of 
$\Delta_b$ on $\epsilon$ is small. Thus, we understand that 
$\Delta_b$ is not playing a very significant role here, and so we 
then proceed to estimate the individual quantities 
that are involved in $\mu_{\tau\tau}$ and $\mu_{bb}$. From 
Eq.~\ref{eq:R1} and Eq.~\ref{eq:R2}, we find that the quantities 
of interest are $\Gamma(h \to gg)$, $\Gamma(h \to WW)$, 
$BR(h \to ZZ)$, $BR(h \to \tau\tau)$ and $BR(h \to bb)$. In 
Fig.~\ref{fig:delb}(b), we display the variation of the ratio of 
partial widths of Higgs to $b\bar b$ and $\tau\tau$ with respect 
to the SM values with the tree level coupling 
$\frac{\sin\alpha}{\cos\beta}$. For most of the 
$2\sigma$ allowed points, we find 25 - 30\% enhancement 
in $bb$ and $\tau\tau$ partial widths 
(see Fig.~\ref{fig:delb}(b)), however 
we check that $\Gamma(h \to ZZ)$ remains 
unaltered. Thus, increase in $\Gamma(h \to bb)$ and thereby increase 
in $\Gamma_{\rm {tot}}$ results in 20-25\% modifications 
in $BR(h \to \tau\tau)$, however, being the dominant decay mode, 
change in $BR(h \to b\bar b)$ is small 
(see Fig.~\ref{fig:delb}(c)). Thus, one can conclude that the 
larger spread in $\mu_{\tau\tau}$ as seen in Fig.~\ref{fig:mu}(b) compared 
to Fig.~\ref{fig:mu}(c) is coming from the interplay 
of the total Higgs decay width and individual Higgs branching 
ratios along with a mild dependence of $\Delta_b$.  

%--------------------------------------------------
%--------------------------------------------------

%%%%%%%%%%%%%%%%%%%%%%%%%%%%%%%%%%%%%%%%%%%%%%%%%%%%%%%%%%%%%%%%%%%%
%%%%% Direct searches ....
%%%%%%%%%%%%%%%%%%%%%%%%%%%%%%%%%%%%%%%%%%%%%%%%%%%%%%%%%%%%%%%%%%%%
%-----------------------------------------------------------------
%\section{MSSM heavy Higgses: Direct searches and Future limits}
\section{Bounds on MSSM heavy Higgses from direct searches}
\label{sec3}

In the previous section, we first describe the global fit analysis, 
and then show that the signal strength measurements of 
the SM-like Higgs boson with different possible final 
state signatures do not exclude the possibility of having additional 
light Higgses (say $\le 600 ~{\rm GeV}$) with moderate values 
of Higgs mixing angle $\alpha$. One can now ask whether these light 
MSSM Higgses are still allowed satisfying the direct search 
bounds at the LHC-8, which is precisely the goal of this section. 
Here we will impose the bounds set by the ATLAS and CMS collaborations on 
the masses and branching ratios of the neutral and charged Higgs 
bosons at the end of 8 TeV run of LHC. 
% Note that, we do not 
% attempt to combine the 7 and 8 TeV data, rather in our analysis 
% we consider the LHC-8 data only. We expect that inclusion of 
% LHC-7 data will not change our results significantly. 
We  calculate the production cross section of 
the neutral heavy Higgses ($H$ and $A$) using 
SuShi (version 1.4.1) \cite{Harlander:2012pb} and charged Higgs ($H^{\pm}$) 
using PYTHIA (version 6.4.28) \cite{Sjostrand:2006za}. The branching ratios of both 
the charged and neutral Higgses are evaluated using 
HDECAY (version 6.41) \cite{Djouadi:1997yw}.     

%------------------------------------------------
\subsection{Neutral Higgs boson searches}
\label{sec:cpeven}

%\vskip 0.5cm 

\subsubsection{Search for $H$ with $\gamma\gamma$ final states}
\label{sec:Hgamgam_8}

\begin{figure}[!htb]
 \begin{center}
 \includegraphics[angle =0, width=0.48\textwidth]{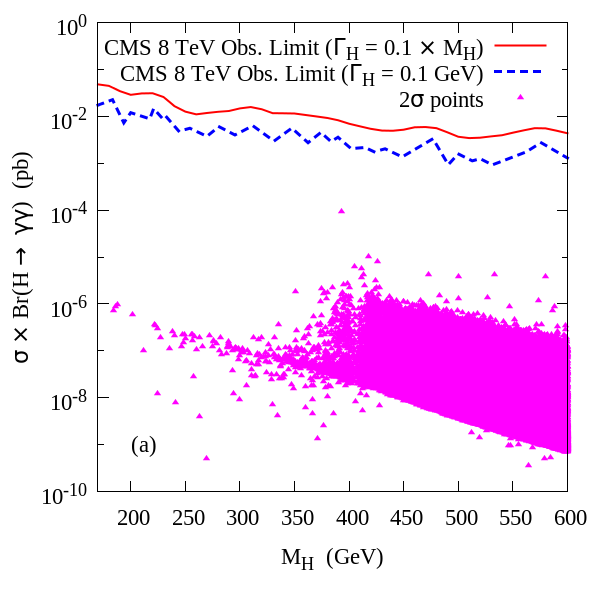}
 \includegraphics[angle =0, width=0.48\textwidth]{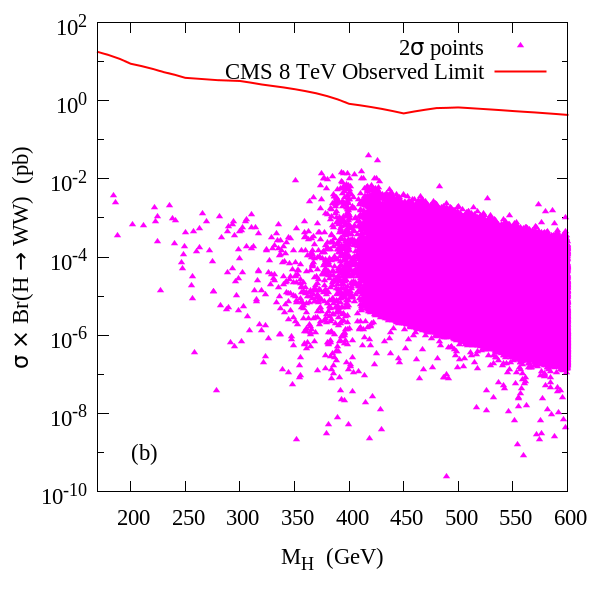}
 \caption{ {\it 
 (a) Left: Scatter plot in  
$M_{H} - [\sigma \times {\rm Br (H \to \gamma\gamma)}]$ plane assuming 
gluon-gluon fusion production process. The solid red (blue dashed) line 
represents the observed upper limits on $\sigma \times {\rm Br (H \to \gamma\gamma)}$ 
at 95\% C.L. by the  CMS collaboration \cite{CMS:2014onr} using LHC-8 data with 
$\Gamma_{H}$ = 0.1 $\times$ $M_H$ (0.1 GeV). 
Magenta coloured triangle shaped points represent 2$\sigma$ allowed parameter 
space from global fits satisfying flavour constraints (see Fig~\ref{fig:ma_tb1}).
(b) Right: Scatter plot in 
$M_{H} - [\sigma \times {\rm Br (H \to WW)}]$ plane when the H is 
produced via ggF. The red solid line indicates 
the observed upper limits on $\sigma \times {\rm Br (H \to WW)}$ 
at the 95\% C.L. by the CMS collaboration \cite{CMS:2012bea} using LHC-8 data.
 }}
 \label{fig:ggH2gamgam}
 \end{center}
 \end{figure}

The di-photon invariant mass distribution plays an important 
role to discover the 125 GeV Higgs boson at the LHC. However, 
the sensitivity of this channel falls rapidly 
with the increase of the SM Higgs boson mass and 
become vanishingly small beyond 150 GeV 
\cite{CMS:2014onr}. In models with additional 
Higgses, di-photon mode can be a 
useful probe to search for 
these heavy resonances. The CMS collaboration has 
searched for the CP-even heavy MSSM Higgs $H$ using 
the di-photon invariant mass distribution with 
19.7 ${\rm fb^{-1}}$ of data at the 8 TeV run of 
LHC. They assume that the Higgs is produced via 
gluon-gluon fusion process. Both narrow and wide 
width heavy resonances are investigated 
with widths ranging from 0.1 to several GeV 
and Higgs masses varying in the range of 
150 GeV to 850 GeV. No excess over the 
SM background has been found, and thus 95\% C.L. upper 
bounds have been set on the production cross section 
times branching ratio in the above-mentioned Higgs 
mass range. In Fig.~\ref{fig:ggH2gamgam}(a), we show the distribution 
of the quantity 
$\sigma \times {\rm Br (H \to \gamma\gamma)}$ for 
all the points of our scanned data set (i.e., 2$\sigma$ 
allowed points obtained after the global analysis 
with various Higgs signal strengths and flavour 
data, see Sec.~\ref{sec2}). We then superimpose 
the bounds set by the CMS collaboration 
for two different choices of the Higgs decay width. 
The red solid line in Fig.~\ref{fig:ggH2gamgam}(a) 
represents the case where Higgs decay width 
is 10\% of the Higgs boson mass, while the blue 
dashed line represents the same but 
with fixed value of the Higgs decay 
width $\Gamma$ = 0.1 GeV. From the figure, one 
can see that all the points of our scanned data set satisfy the 
CMS bounds. A further investigation reveals that the Higgs 
to $\gamma\gamma$ branching ratio for all the points 
corresponding to the scanned dataset varies between 
$10^{-6} - 10^{-7}$ for all the points, and 
thus makes the quantity $\sigma \times {\rm Br (H \to \gamma\gamma)}$ 
small enough to evade the CMS bound.

%-----------------------------------
\subsubsection{Search for $H$ with $WW$ final states}
\label{sec:HWW_8}

In the SM, the decay of Higgs boson to electro-weak 
gauge bosons ($W/Z$) provides interesting signatures at the LHC. 
%For example, the di-photon mode has the best sensitivity 
%to discover the Higgs boson, 
In fact, the $ZZ$ mode has a very good sensitivity 
to precisely measure the Higgs boson mass and its 
spin and parity \cite{Aad:2014aba,Khachatryan:2014jba}. 
The CMS collaboration has searched for a SM-like 
Higgs boson decaying to a pair of $\rm W$ bosons 
with (5 + 19.3) ${\rm fb^{-1}}$ data collected 
at $\sqrt s$ = (7 + 8) TeV at the LHC \cite{CMS:2012bea}. 
Analyzing the data with lepton, jets plus missing transverse 
energy final state signature, CMS collaboration excludes SM-like 
Higgs bosons in the mass ranges 170-180 GeV and 
230-545 GeV at 95\% C.L.. One can translate the bounds 
on the cross sections obtained from search of SM-like 
Higgs boson via this channel to the same corresponding to 
a model with additional Higgses. We calculate the product 
of production cross section and $WW$ branching ratio, 
$\sigma \times {\rm Br (H \to WW)}$, for our scanned points 
and then compare our findings with the CMS data. 
Note that we do not combine the 7 and 8 TeV data of LHC, 
rather we consider the exclusion limits of the 
8 TeV data only. In Fig.~\ref{fig:ggH2gamgam}(b), we superimpose 
the CMS exclusion limit on the 2$\sigma$ allowed points 
obtained from global fit analysis. We find that 
updated CMS bound on production cross section with Higgs decaying 
to pair of $W$s is not sensitive enough to 
exclude the scanned parameter space. As we have already discussed, 
for almost all the points we are near the alignment 
limit (i.e. $(\beta - \alpha) \sim \frac{\pi}{2}$), which 
results into highly suppressed branching ratio of 
Higgs to the $WW$ mode, $\rm Br(H \to WW) \sim 10^{-2} - 10^{-4}$. 
This is the reason why the quantity 
$\sigma \times {\rm Br}(H \to WW)$ is very small, and thus 
become insensitive to the CMS bound. We also 
find that even if we consider the (7+8) TeV combined CMS 
exclusion limit the parameter space will still remain 
allowed by the LHC data.    

%-----------------------------------
\subsubsection{Search for $H$ with $hh$ ($b\bar b b \bar b $ and 
$b\bar b \gamma\gamma$) final states}
\label{sec:Hhh_8}

\begin{figure}[!htb]
 \begin{center}
 \includegraphics[angle =0, width=0.48\textwidth]{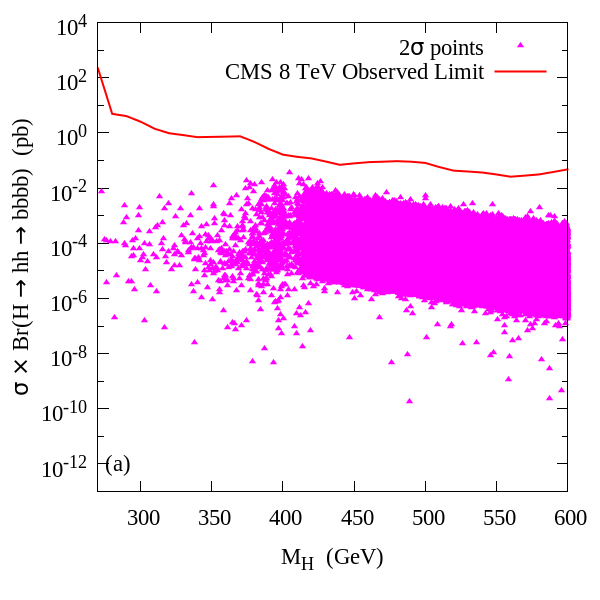}
 \includegraphics[angle =0, width=0.48\textwidth]{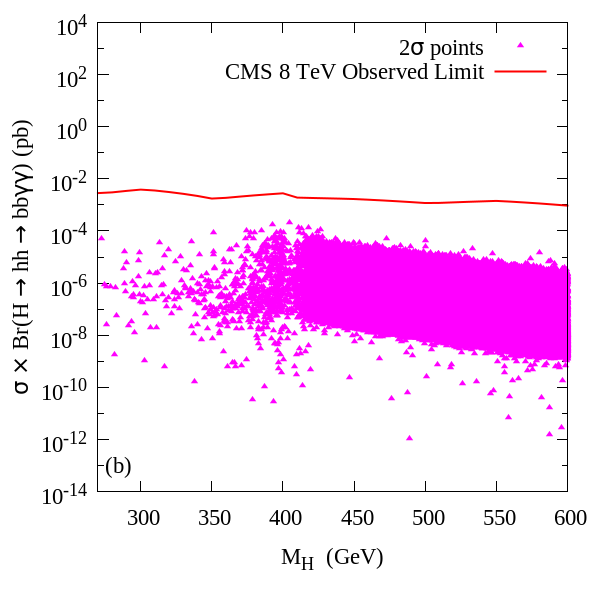}
 \caption{ {\it 
(a) Left: Scatter plot in  
$M_{H} - [\sigma \times {\rm Br (H \to h h \to b \bar b b \bar b)}]$ plane 
assuming gluon-gluon fusion production process for 2$\sigma$ allowed 
parameter space (represented by magenta points). The solid red line 
represents the observed upper limits on $\sigma \times {\rm Br (H \to h h \to b \bar b b \bar b)}$ 
at 95\% C.L. by the CMS collaboration \cite{CMS:2014eda} using 8 TeV data. 
(b) Right: 
Scatter plot in 
$M_{H} - [\sigma \times {\rm Br (H \to h h \to b \bar b \gamma\gamma)}]$ 
for ggF production mode. 
The red solid line indicates the observed upper limits on
 $\sigma \times  {\rm Br (H \to h h \to b \bar b \gamma\gamma)}$ 
at the 95\% C.L. by the CMS collaboration \cite{CMS:2014ipa} using LHC 8 TeV data. 
 }}
 \label{fig:GGH2hh}
 \end{center}
 \end{figure}

The Higgs pair production cross section 
in the SM is very small, around 10 {\rm fb} at 
$\sqrt s = $ 8 TeV \cite{CMS:2014ipa}. 
However, well-motivated 
BSM physics models predict the decay of 
narrow-width heavy resonances to a pair of 125 GeV 
Higgses, and thus one can expect enhancement 
in the 125 GeV Higgs pair production cross section.
Search for these heavy resonances decaying to a pair 
of h, i.e. $ p p \to X \to hh $ with X being 
the new heavy resonance, is performed 
by the CMS collaboration in the mass range of 
260 - 1100 GeV with 19.7 $\rm fb^{-1}$ of data at 
$\sqrt s$ = 8 TeV \cite{CMS:2014ipa,CMS:2014eda}
The final state signatures that have 
been investigated by the CMS collaboration are: 
($i$) $b\bar b b \bar b $ where both the Higgses decaying 
to $b\bar b$, and ($ii$) $b\bar b \gamma\gamma$ where one of 
the Higgs decays to $b\bar b$ while other decays to a pair of 
photons. The reconstruction of heavy resonance is possible 
in both the above-mentioned signatures, however 
4$b$ final state is experimentally challenging while 
$b\bar b \gamma\gamma$ channel has less background 
contamination and very good di-photon mass resolution. 
The observations by the CMS collaboration 
are consistent with the SM, and so 95\% exclusion limits 
on the production cross sections are placed for the 
entire mass range of the heavy resonance. At the LHC, the 
MSSM heavy CP even Higgs (H) can be generously 
produced from the ggF process, which then decays 
to a pair of 125 GeV Higgses. So, the bounds 
on the production cross section of new heavy resonances can be 
directly used to probe/exclude certain mass range of 
the heavy Higgs boson. To do so, we 
calculate the quantities 
$\sigma(p p \to H) \times {\rm Br (H \to h h \to b \bar b b \bar b)}$ and 
$\sigma(p p \to H) \times {\rm Br (H \to h h \to b \bar b \gamma\gamma)}$ 
for all our scanned data set and in Fig.~\ref{fig:GGH2hh}(a) and 
Fig.~\ref{fig:GGH2hh}(b), we show the distributions of 
those two quantities respectively. The 95\% C.L. upper 
limits (solid red line) imposed by the CMS collaboration 
on the respective distributions are also superimposed.

%----------------------------------------------
%----------------------------------------------

The branching fraction of the MSSM heavy Higgs H to hh becomes 
sizable only for small $\tan\beta$ ($\le 5$) and 
low $M_A$ \cite{Djouadi:2005gj}. Once $M_A$ 
becomes $\ge$ 350 GeV, the $t\bar t$ decay mode opens 
up and dominates in the rest of the parameter space. We 
find that so far CMS data can not 
exclude our 2$\sigma$ allowed parameter space, however the 
regions close to CMS exclusion lines corresponds to 
smaller values of $\tan\beta$, and so with 
one/two orders of improved measurement of di-Higgs production 
cross section at the Run-II of LHC, one can probe/exclude such 
regions of the parameter space. The ATLAS collaboration has 
also searched for the heavy resonances 
(Kaluza-Klein excitation mode graviton)
decaying to a pair of SM Higgses with both the 
Higgs decaying to a pair of b-quarks in the context of the
Randall-Sundrum model \cite{ATLAS-CONF-2014-005}. The 
95\% exclusion limits set by the ATLAS collaboration are 
sensitive for heavy resonances (mass $>$ 500 GeV) only, 
and thus is not directly 
applicable to the parameter space of interest, so we 
do not consider them in our analysis.  
%----------------------------------------------

\subsubsection{Search for $H/{A}$ with $\tau^{+}\tau^{-}$ final states}
\label{sec:Htautau_8}

\begin{figure}[!htb]
\begin{center}
{\includegraphics[angle =0, width=0.48\textwidth]{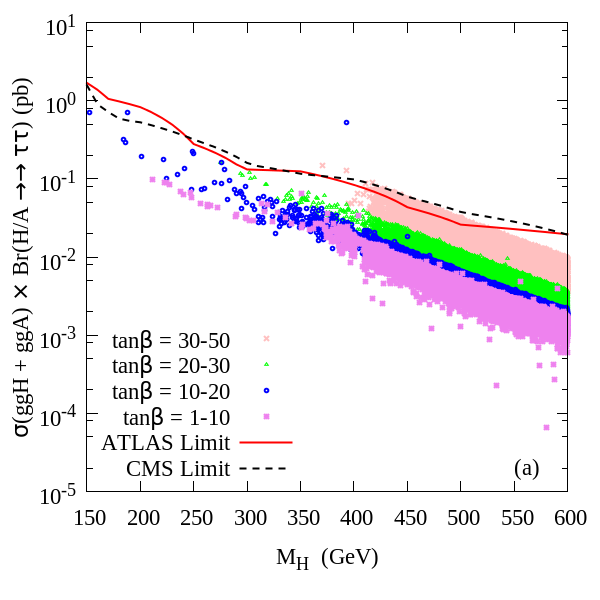} }
{\includegraphics[angle =0, width=0.48\textwidth]{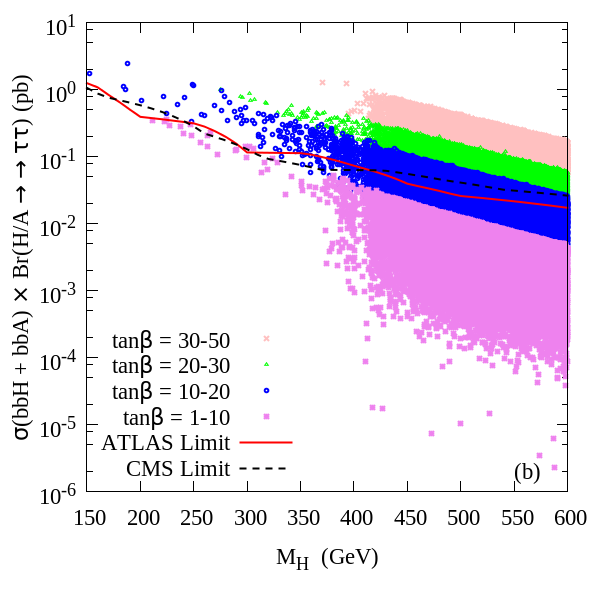} }
\caption{ {\it 
Scatter plot in  
$M_{H} - [\sigma \times {\rm Br (\Phi \to \tau^+ \tau^-)}]$ plane 
where $\Phi =$ H, A produced via (a) gluon fusion and (b) in association with
$b$-quarks.
The solid red (black dashed) line  
represents the observed upper limits on $\sigma \times {\rm Br (\Phi \to \tau^{+}\tau^{-})}$ 
at 95\% C.L. by the ATLAS (CMS) collaboration using LHC-8 data \cite{Aad:2014vgg,Khachatryan:2014wca}. 
Combined contributions of $\sigma \times {\rm {Br(\Phi \to \tau^+ \tau^-)}}$  
of both $H$ and $A$ for the 
2$\sigma$ allowed parameter space with $30 < \tan\beta < 50$, 
$20 < \tan\beta < 30$, $10 < \tan\beta < 20$ and $1 < \tan\beta < 10$ 
are shown in pink (cross), green (triangle), blue (circle) 
and violet (square) respectively. 
 }}
\label{fig:Htautau}
\end{center}
\end{figure}

The coupling of the MSSM heavy Higgs ($H$) and pseudoscalar Higgs ($A$) 
with down type fermion ${f_d}$ (say bottom quark, tau lepton) is proportional 
to $\cos\alpha/\cos\beta$ and $\tan\beta$ respectively. So, for a 
fixed value of Higgs mixing angle $\alpha$, both the 
couplings $H{f_d}\bar {f_d}$ and $A{f_d}\bar {f_d}$ increases with 
$\tan\beta$.   
Thus for large values of $\tan\beta$ (say $\ge$ 10), 
both $H$ and $A$ dominantly decays to $b\bar b$ ($\sim$ 90\%) and 
$\tau^{+}\tau^{-}$ ($\sim$ 10\%), resulting into strong suppression 
in all other decay modes \cite{Djouadi:2005gj}. 
The production of the MSSM 
heavy Higgses is also primarily controlled by $\tan\beta$. The Higges 
are dominantly produced via gluon-gluon fusion process and associated 
production with b-quarks i.e., $b\bar b\Phi$ ($\Phi$= H,A). 
All other production processes like 
VBF, associated production with gauge bosons, associated production 
with top quarks are dominantly suppressed for the heavy Higgses H/A 
due to small $HVV$/$AVV$ and $Ht\bar t$/$At\bar t$ couplings 
\cite{Djouadi:2005gj}. The 
ATLAS and CMS collaborations at the LHC have studied the signatures 
of these heavy Higgses ($H/A$) produced via ggF and b-quark 
associated production processes, and decays to pair of $\tau$-leptons 
\cite{Aad:2014vgg,Khachatryan:2014wca}. No excesses are observed over the SM backgrounds, 
and thus model independent bounds are placed on the production 
cross section times branching ratio 
$\sigma \times {\rm {Br(\Phi \to \tau^+ \tau^-)}}$ for different values of 
$M_{\Phi}$ with $\Phi$ = H/A. In Fig.~\ref{fig:Htautau}(a) and 
Fig.~\ref{fig:Htautau}(b), we display our scanned points 
along with the upper limit on production cross section 
by the ATLAS and CMS collaborations for two dominant 
production modes, left panel corresponds to the ggF process 
while right panel represents 
the associated production process $b\bar b\Phi$, $\Phi = H,A$. 
In order to compare our findings with the ATLAS \cite{Aad:2014vgg}
 and CMS \cite{Khachatryan:2014wca} data, in 
Fig.~\ref{fig:Htautau}(a) and Fig.~\ref{fig:Htautau}(b) we show 
the distributions of 
$\sigma \times {\rm {Br(\Phi \to \tau^+ \tau^-)}}$ combining the 
contributions of both $H$ and $A$ for the two different production 
mechanism.

%--------------------------------------------------
%--------------------------------------------------

It is evident from Fig.~\ref{fig:Htautau} that the 
LHC data from the $H\to \tau^+ \tau^-$ channel has  
significant impact on the 2$\sigma$ allowed 
parameter space. In fact, the 
entire region with $\tan\beta >$~20 and significant amount 
of parameter space with $\tan\beta >$~10 are excluded when 
the Higgs is produced via $b\bar b \Phi$ process. However, 
sensitivity of this channel with ggF production process is 
negligible. For a better understanding of the sensitivity 
of $\tau^+ \tau^-$ channel, we display the scanned dataset with 
explicit $\tan\beta$ dependence, regions with $30 < \tan\beta < 50$, 
$20 < \tan\beta < 30$, $10 < \tan\beta < 20$ and $1 < \tan\beta < 10$ 
are shown in pink (cross), green (triangle), blue (circle) 
and violet (square) respectively. One 
can now clearly see that the regions with large 
$\tan\beta$ implies large coupling with $b$ and $\tau$s and so 
large $\sigma \times {\rm Br}$, and thus more stringent 
constraint by the LHC data.

%--------------------------------------------------
\subsubsection{Search for $A$ with $Zh$ final states}
\label{sec:AZh_8}

%--------------------------------------------------
\begin{figure}[!htb]
\begin{center}
{\includegraphics[angle =0, width=0.32\textwidth]{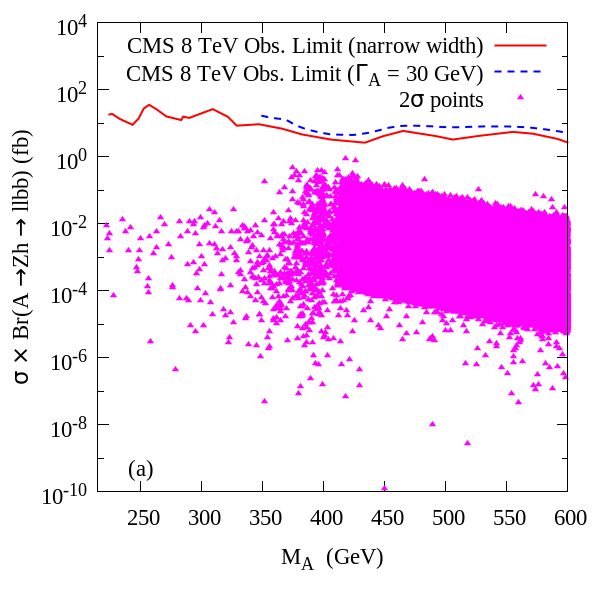} }
{\includegraphics[angle =0, width=0.32\textwidth]{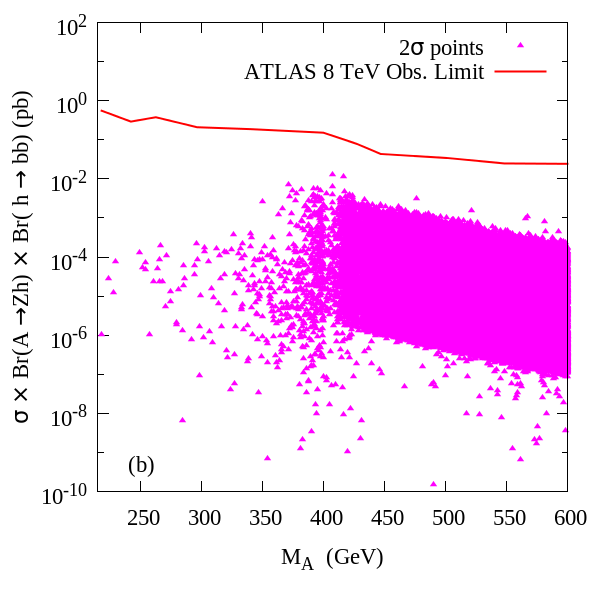} }
{\includegraphics[angle =0, width=0.32\textwidth]{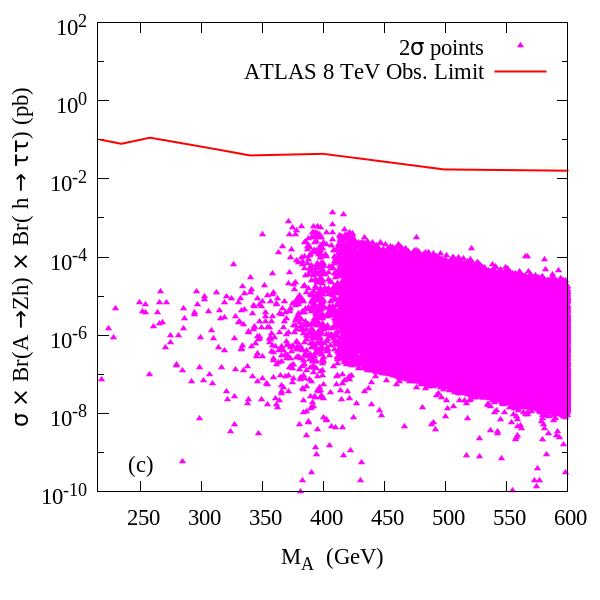} }
\caption{ {\it 
(a) Left: Scatter plot in  
$M_{A} - [\sigma \times {\rm Br (A \to Z h \to \ell ^+ \ell^-  b \bar b)}]$ plane 
assuming gluon-gluon fusion production process for 2$\sigma$ allowed 
parameter space (represented by magenta points). 
The solid red (blue dashed) line 
represents the observed upper limits on 
$\sigma \times {\rm Br (A \to Z h \to \ell^+ \ell^-  b \bar b)}$ 
at 95\% C.L. by the  CMS collaboration using LHC-8 data 
\cite{CMS:2014yra} with 
narrow width ($\Gamma_{A}$ = 30 GeV) assumption.
(b)~Middle: 
Scatter plot in  
$M_{A} - [\sigma \times {\rm Br (A \to Z h)} \times{\rm Br(h \to b \bar b)}]$ plane 
for ggF production mode. 
The red solid line indicates the observed upper limits on
$\sigma \times {\rm Br (A \to Z h)} \times {\rm Br (h \to b \bar b)}$
at the 95\% C.L. by the ATLAS collaboration using LHC 8 TeV data \cite{Aad:2015wra}. 
(c)~Right: 
Scatter plot in  
$M_{A} - [\sigma \times {\rm Br (A \to Z h)} \times{\rm Br(h \to \tau^+ \tau^-)}]$ plane 
for ggF production mode. 
The red solid line shows the observed upper limits on
$\sigma \times {\rm Br (A \to Z h)} \times {\rm Br(h \to \tau^+ \tau^-)}$
at the 95\% C.L. by the ATLAS collaboration using LHC 8 TeV data \cite{Aad:2015wra}. 
}}
\label{fig:cmsAzh}
\end{center}
\end{figure}
%--------------------------------------------------
 
Similar to the case where we discussed the possibility 
of heavy resonances decaying to a pair of MSSM lightest 
Higgs bosons (see Sec.~\ref{sec:Hhh_8}), one 
can also consider the possibility of having $h$ in 
association with $Z$ boson from the decay of pseudoscalar 
Higgs boson $A$ when $M_A$ is greater than 
($M_{h} + M_{Z}$). However, the decay rate 
$\Gamma(A \to Z h)$ is appreciable only at very low 
values of $\tan\beta$ ($<$ 10) and for $M_{A}$ below 
the $t\bar t$ threshold i.e., approx. 350 GeV. Once the 
$t\bar t$ decay opens up, it dominates for all values 
$M_{A} >$ 350 GeV and $\tan\beta < 10$ \cite{Djouadi:2005gj}. 
The CMS collaboration at the LHC has searched for the heavy 
pseudoscalar Higgs bosons when it decay to a $Z$ boson and a 
light Higgs boson ($h$). The final state includes two 
opposite sign leptons (from $Z$ decay) and two b-quarks 
(from $h$ decay). Thus, one can fully reconstruct the mass 
of $A$ using the four momentum information of the 
final state leptons and b-jets. In fact three clear 
distinguishable resonance peaks, at $M_Z$, $M_h$ and $M_A$, 
are expected for the signal events. Both the ATLAS  and CMS 
collaborations have performed the search for the heavy 
pseudoscalar Higgs bosons at the 8 TeV run of LHC 
\cite{Aad:2015wra,CMS:2014yra}. No significant excesses over SM 
background are observed in the ATLAS and CMS data. 
Thus model independent 95\% C.L. upper 
limits are imposed on the production cross section of $A$ 
times ${\rm Br (A \to Z h)}$. In Fig.~\ref{fig:cmsAzh}(a) and (b), 
we present the $2\sigma$ allowed scanned data points along with 
the CMS and ATLAS exclusion limits respectively. The solid red  
exclusion line in Fig.~\ref{fig:cmsAzh}(a) assume narrow-width 
resonance, while the blue dashed line consider the 30 GeV 
decay width of $A$ boson. The ATLAS collaboration has also performed 
their study with narrow width approximation, however they have 
analyzed two possible decay modes of the Higgs boson, namely 
$b\bar b$ and $\tau^{+}\tau^{-}$. In Fig.~\ref{fig:cmsAzh}(b) 
we display the allowed parameter space with $b\bar b$ final state, 
while Fig.~\ref{fig:cmsAzh}(c) shows the same but with 
$\tau^{+}\tau^{-}$ final state. As we already mentioned, this 
decay mode plays an important role only in the region with 
small $\tan\beta$ and low $M_A$. We find that in the parameter 
space of our interest, ATLAS and CMS data are not sensitive enough to 
impose any additional constraints. We expect these heavy 
pseudoscalar Higgses will be probed with improved 
measurement of these decay modes in the run-II of LHC.

\subsection{Charged Higgs boson searches}

\subsubsection{Search for $H^{\pm}$ with $\tau\nu$ and $c\bar s$ final states}
\label{sec:CHtaunu_8}

The SM particle content does not include a 
charged scalar particle, however models 
with additional Higgs doublets predict 
charged Higgs bosons ($H^{\pm}$). Thus the 
discovery of a charged scalar particle is a 
clear signature of the BSM physics. The large 
electron positron collider (LEP) searched for the 
charged Higgs bosons with center-of-mass energy 
$\sqrt s$ = 209 GeV. LEP did not found any signal of 
charged Higgs boson which then leads to put 
bounds on mass of the charged Higgs boson 
$M_{H^\pm} > $ 78.6 GeV \cite{Searches:2001ac}. 
The production and decay of $H^{\pm}$ primarily depend 
on $M_{H^{\pm}}$ and top quark mass, more precisely 
on whether $M_{H^{\pm}} < M_{\rm top}$ or 
$M_{H^{\pm}} > M_{\rm top}$. If the charged Higgs 
boson is lighter than the top quark, i.e., 
$M_{H^\pm} < ({m_{t} - m_{b}})$, 
then $H^\pm$ is mostly produced from the $t\bar t$ 
process. The decay 
of $H^{\pm}$ depends on the coupling of the charged Higgs boson 
with the fermions which is mainly controlled by $\tan\beta$. 
The coupling is large for very low and very large 
values of $\tan\beta$ and small for intermediate values. 
Thus, for light enough $H^{\pm}$ it is primarily 
produced from the $t\bar t$ process and dominantly 
decays into $\tau\nu_{\tau}$ final states. It is to be noted 
that light charged Higgs bosons 
($M_{H^\pm} < M_{t}$) can also decay to 
a charm and a anti-strange quark, and thus in certain 
regions of parameter space one can expect some competition 
between the $c\bar s$ and $\tau\nu_{\tau}$ decay modes. Now, 
if the $H^{\pm}$ is heavy i.e., $M_{H^{\pm}} > M_{\rm top}$, 
then they are mainly produced in associated production 
with a top quark i.e., $p p \to t H^{\pm} + X$. For 
small $\tan\beta$, $H^{\pm}$ exclusively decays to a top 
and bottom quark, however for large values of $\tan\beta$ 
the decay of $H^{\pm}$ to $\tau\nu_{\tau}$ is not 
negligible i.e., ${\rm Br}(H^{\pm}\to \tau^{\pm}\nu_{\tau})\sim$ 
10\% \cite{Djouadi:2005gj}.  

%Top 
%quark decay to $bH^\pm$ be as large as 10 -20\% for 
%very small ($<$ 3) or very large ($>$ 30) values of $\tan\beta$. 

Both CMS and ATLAS collaborations have searched for 
the charged Higgs bosons at the 8 TeV run of LHC using 
the top-quark pair production process and associated 
production of $H^{\pm}$ with a top quark. Search for 
the light charged Higgs bosons decaying to a $\tau$ and a 
tau-neutrino ($\nu_{\tau}$), and/or to a charm and 
strange quark are also presented by the ATLAS and CMS 
collaborations \cite{Aad:2014kga,CMS:2014cdp,
Aad:2013hla,CMS:2014kga}. The results seem to agree with the 
SM predictions and non observation of the excesses 
leads to 95\% C.L. exclusion limits on the 
production cross section times branching ratios 
for different values of $M_{H^\pm}$. 
%The CMS collaboration at the LHC
%has also searched for charged Higgses when they are produced from the
%top pair production process with one top decaying to a bottom quark
%and a $H^\pm$, with $H^+\to c \bar{s}$. The looked for the signal
%with final state signature containing one isolated muon (coming from the
%top quark decay), multiple jets ($\ge$ 4) and missing transverse
%energy. No significant deviation from the SM expectations is observed,
%and so model-independent upper limits are set on ${\rm BR}(t \to H^{+}b)$
%assuming ${\cal B}(H^{+} \to c\bar{s}) = 100\%$ for $M_{H^\pm}$
%between 90 to 160 GeV. 

\begin{figure}[!htb]
\begin{center}
{\includegraphics[angle =0, width=0.48\textwidth]{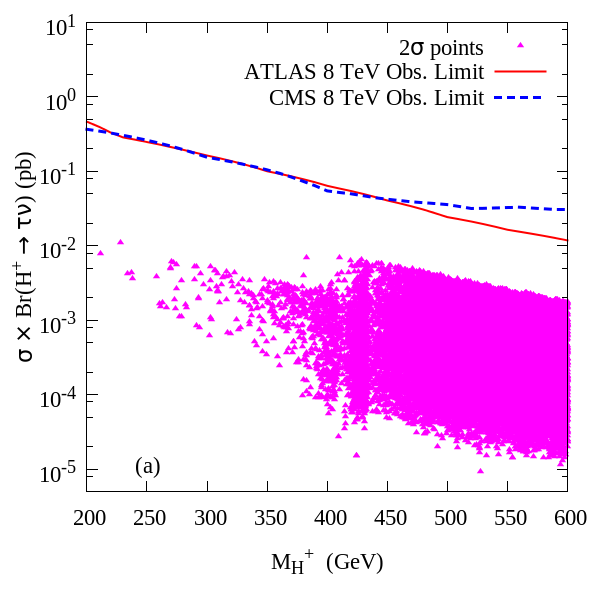}} 
{\includegraphics[angle =0, width=0.48\textwidth]{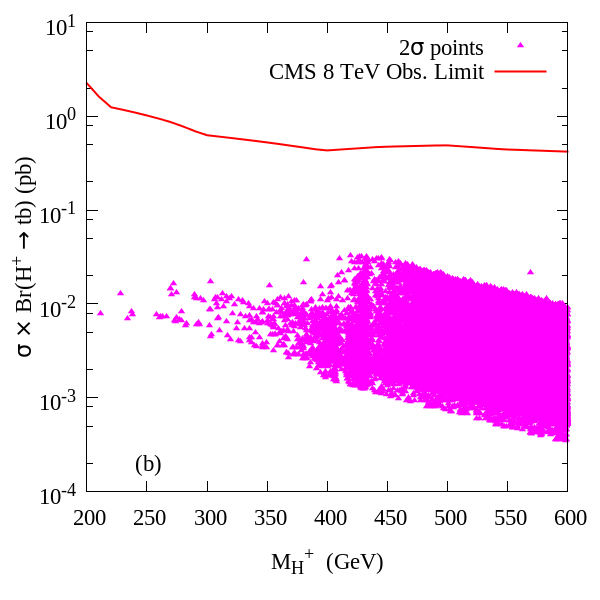}} 
\caption{ {\it 
(a) Left: Scatter plot in  
$M_{H^\pm} - [\sigma \times {\rm Br (H^\pm \to \tau^\pm \nu_{\tau})}]$ plane 
for ggF production. 
Magenta points represent 2$\sigma$  allowed regions. 
The solid red (blue dashed) line represents the observed upper limits on 
$\sigma \times {\rm Br (H^\pm \to \tau^\pm \nu_{\tau})}$
at 95\% C.L. by the ATLAS \cite{Aad:2014kga} (CMS ) collaboration using LHC 8 TeV data.
(b)~Right: Scatter plot in  
$M_{H^\pm} - [\sigma \times {\rm Br (H^\pm \to t \bar b)}]$ plane 
for ggF production. 
The solid red line shows the observed  upper limits on 
$\sigma \times {\rm Br (H^\pm \to t \bar b)} $
at 95\% C.L. by the CMS collaboration using LHC 8 TeV data.}}
\label{fig:hpltaunu}
\end{center}
\end{figure}

Now, we find that all the points of 
the scanned data set corresponds to $M_{H^\pm} >$ 200 GeV and thereby 
the decay $t \to b H^{\pm}$ is kinematically forbidden, and 
so ATLAS and CMS bounds on $M_{H^\pm}$ using the $t\bar t$ sample 
have no effect on the parameter space of interest. However, 
we do impose the ATLAS and CMS bounds on $M_{H^{\pm}}$ considering 
the production of charged Higgs boson in association to a top and bottom 
quark with $H^{\pm}$ decaying to $\tau^\pm\nu_{\tau}$. 
In Fig.\ref{fig:hpltaunu}(a), we superimpose the ATLAS and 
CMS bounds on the points corresponding to our data set. We check 
that the bounds using the $c\bar s$ mode are not 
applicable. From Fig.\ref{fig:hpltaunu}(a) one can infer that 
one/two orders of improvement in cross section 
measurement might help to probe the parameter 
space of interest.

%--------------------------------------------------
%--------------------------------------------------

\subsubsection{Search for $H^{\pm}$ with $t\bar b$ final states}
\label{sec:CHtb_8}
 
As we have already mentioned, the decay of $H^{\pm}$ to a top 
and bottom quark dominates in the regions with 
$M_{H^\pm} > M_{top}$. They are primarily produced via 
$g g \to tbH^{\pm}$ process. However processes like 
$q\bar q \to H^{+}H^{-}$, associated production with neutral 
Higgses can give small contributions to the dominant $tbH^{\pm}$ 
process. The decay of $H^{\pm}$ is completely controlled 
by the single parameter $\tan\beta$. For small values of $\tan\beta$ 
it is the dominant decay mode (branching ratio is close to 
unity), however at high $\tan\beta$ the branching ratio 
decreases slightly ${\rm Br}(H^{\pm} \to t \bar b) \sim $ 90\% 
while ${\rm Br}(H^{\pm} \to \tau^\pm \nu_{\tau}) \sim $ 10\%. 
The CMS collaboration have searched for the charged 
Higgs bosons with $H^\pm$ decaying to a top 
and bottom quark \cite{CMS:2014pea}. The 
process under consideration looks like 
$gg \to H^{+}tb \to (\ell \nu_{\ell} b b) (\ell^{\prime} \nu_{\ell^\prime} b) b$,
with $\ell$, $\ell^{\prime}$ being an electron or a muon. The 
search is performed with 19.7 $\rm fb^{-1}$ of data at $\sqrt s$ = 8 TeV. 
No evidence for a charged Higgs signal is found and thus upper 
limits on the production rates are placed for $H^{\pm}$ masses in the 
range of 180-600 GeV. We calculate the $g g \to tbH^{\pm}$ 
production cross section for all our valid points, and then 
impose the CMS bounds. We present our results in 
Fig.~\ref{fig:hpltaunu}(b), where the solid red line indicates 
the CMS exclusion limit. It is evident from the figure that $t\bar b$ 
mode is not sensitive enough to put any strong bound on the parameter 
space.

%--------------------------------------------------
%%%%%%%%%%%%%%%%%%%%%%%%%%%%%%%%%%%%%%%%%%%%%%%%%%%%%%%%%%%%%%%%%%%%%%%%%%%%%%%%%%%%%%%
%
%......... FUTURE LIMITS ...
%
%%%%%%%%%%%%%%%%%%%%%%%%%%%%%%%%%%%%%%%%%%%%%%%%%%%%%%%%%%%%%%%%%%%%%%%%%%%%%
\section{SUSY Higgs : Future limits }
\label{sec4}

%-----------------
\subsection{Heavy Higgs search}

In the last section, we discussed the limits on 
the allowed parameter space of the MSSM Higgs sector 
coming from the direct searches at LHC run-I. We show that 
significant amount of parameter space with additional light 
Higgses are still allowed by the LHC data. Additionally, we also 
found that the production of heavy Higgs bosons in association with 
the bottom quarks with $H/A$ decaying to $\tau^{+}\tau^{-}$ 
provides the most stringent bound on our parameter space 
and the regions with large values of $\tan\beta$ ($>$ 20) 
are excluded by the $\tau^{+}\tau^{-}$ decay mode. In 
this section, we discuss the sensitivity 
of HL-LHC for the heavy Higgses through the following decay 
modes: $H \to hh$, $H \to t\bar t$, $A \to Z h$ and 
$H/{A} \to \tau^{+}\tau^{-}$. Note that, these 
channels\footnote{Reach of heavy Higgs bosons have been discussed 
in the literature in the context of 8/14 TeV run of LHC, for details see 
Refs.~\cite{Carena:2013qia,Altmannshofer:2012ks,Carena:2014nza 
,slimsusy, pilaftis}.} 
are expected to be most sensitive at the regions 
with $\tan\beta <$ 20. Additionally, we also consider the 
decay $ H \to ZZ$ following the ATLAS and CMS analysis 
\cite{ATL-PHYS-PUB-2013-016, CMS:2013dga} at 
14 TeV run of LHC with 3000 $\rm fb^{-1}$ of integrated 
luminosity.

\subsubsection{Search for $H$ with $4\ell$ final states}
\label{sec:HZZ_14}

%--------------------------------------------------
\begin{figure}[!htb]
\begin{center}
{\includegraphics[angle =0, width=0.6\textwidth]{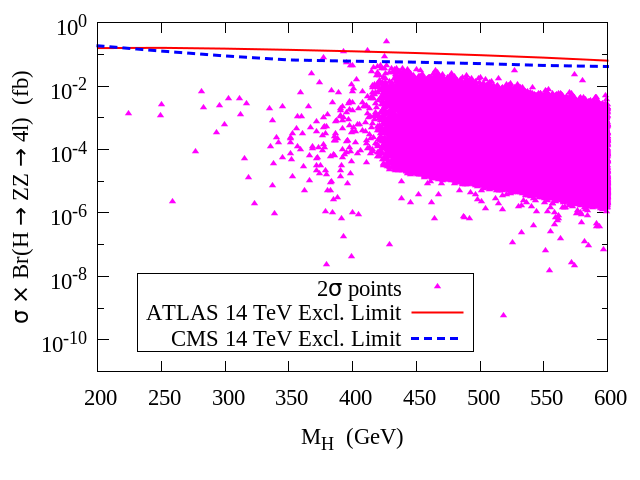} }
\caption{ {\it 
Scatter plot in  
$M_{H} - [\sigma \times {\rm Br (H \to ZZ \to \ell^+ \ell^- \ell^+ \ell^- )}]$ 
plane assuming gluon-gluon fusion production process.
Magenta colored points represents for 2$\sigma$ allowed 
parameter space. The solid red line (blue dashed line)  
represents the expected upper limits on 
$\sigma \times {\rm Br (H \to ZZ \to \ell^+ \ell^- \ell^+ \ell^- )}$ 
at 95\% C.L. by the ATLAS (CMS) collaboration \cite{ATL-PHYS-PUB-2013-016, CMS:2013dga}  for 
14 TeV LHC with $\lum = $ 3000~$\ifb$. 
 }}
\label{fig:HZZ_14}
\end{center}
\end{figure}
%--------------------------------------------------

Both the ATLAS and CMS collaborations have 
looked for the heavy Higgs bosons which are 
produced via ggF process and decayed 
into $ZZ$ ($Z \to \ell \ell$, $\ell = e, \mu$) 
at 14 TeV HL-LHC 
\cite{ATL-PHYS-PUB-2013-016, CMS:2013dga}. In their analyses, 
the ATLAS collaboration has considered the decay width of 
$H$ to be the same as that of a SM-like Higgs boson 
for a given mass, while the CMS collaboration calculated 
the width of $H$ assuming $\tan\beta$ = 1 and $\cos(\beta -\alpha$) = - 0.06. 
The expected 95\% C.L. exclusion limits on 
$\sigma \times$ Br($H \to ZZ\to 4\ell)$ at $\lum$ = 3000~$\ifb$ 
by the ATLAS and CMS collaborations are 
quite similar as shown in Fig.~\ref{fig:HZZ_14} by 
red solid and blue dashed line respectively. We calculate the 
quantity $\sigma \times$ Br($H \to ZZ\to 4\ell)$ for all the 
points in our scanned data set and then compare our results 
with ATLAS and CMS predictions\cite{ATL-PHYS-PUB-2013-016, CMS:2013dga}.
 In Fig.~\ref{fig:HZZ_14} we overlay 
the experimental predictions with our scanned data set shown 
in magenta triangles. We find that for most 
of the parameter space points, even for the points with large 
production cross sections, the branching ratio 
Br($H \to ZZ)$ is very small 
($\sim 10^{-3}$ to $10^{-5}$)\footnote{ The result is consistent with 
the alignment limit.}, 
which results into smaller values of 
$\sigma \times$ Br($H \to ZZ\to 4\ell)$. We thus 
find that most of the parameter space points are beyond the reach 
of HL-LHC for $H \to 4\ell$ final state.  

%expected upper limit on $\sigma \times$ BR($H \to ZZ\to 4\ell)$ for 
%$M_H$ = 200 (600) GeV is 100 (25) times stronger than that expected 
%for a SM-like Higgs boson in the same mass range.  

%%%%%%%%%%%%%%%%%%%%%%%%%%%%%%%%%%%%%%%%%%%%%%
%%%%%%%%%%%%%%%%%%%%%%%%%%%%%%%%%%%%%%%%%%%%%%

\subsubsection{Search for pseudoscalar $A$ with $\ell^+ \ell^- b \bar b$ final states}

%--------------------------------------------------
\begin{figure}[!htb]
\begin{center}
{\includegraphics[angle =0, width=0.6\textwidth]{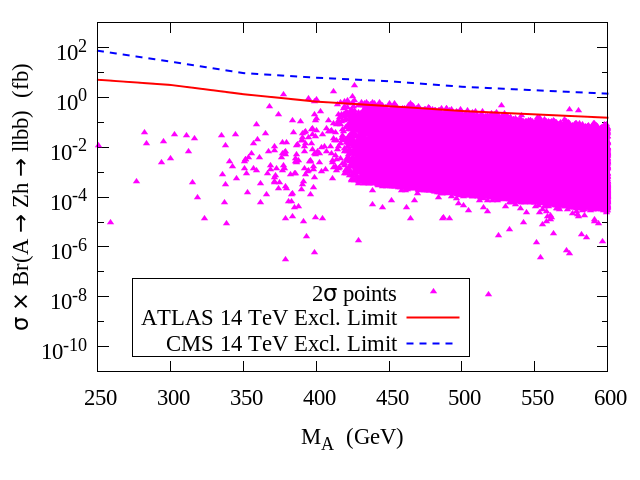}}
\caption{ {\it 
Scatter plot in  
$M_{A} - [\sigma \times {\rm Br (A \to Zh \to \ell^+ \ell^- b \bar b)}]$ 
plane assuming gluon-gluon fusion production process.
Magenta colored points represents for 2$\sigma$ allowed 
parameter space. The solid red line (blue dashed line)  
represents the expected upper limits on 
$\sigma \times {\rm Br (A \to Zb \to \ell^+ \ell^- b \bar b)}$ 
at 95\% C.L. by the ATLAS (CMS) collaboration  for 
14 TeV LHC with $\lum = $ 3000~$\ifb$ \cite{ATL-PHYS-PUB-2013-016, CMS:2013dga}. 
}}
\label{fig:AZH_14}
\end{center}
\end{figure}
%--------------------------------------------------

In the regions below the $t\bar t$ threshold and small 
$\tan\beta$ the pseudoscalar Higgs $A$ decays to $Zh$ with an 
appreciable amount. An interesting feature 
of this decay mode with $Z \to \ell^{+} \ell^{-}$ and 
$h \to b \bar b$ is that one can fully reconstruct the 
mass $A$ using the four momentum of the leptons 
and b-jets. ATLAS and CMS collaborations 
have analysed the sensitivity of this channel at the 
HL-LHC via ggF process \cite{ATL-PHYS-PUB-2013-016, CMS:2013dga}. 
In order to present the expected 95\% C.L. exclusion limits ATLAS has 
assumed a narrow width approximation (i.e., width 
of $A$ is much smaller than the experimental 
resolution) while CMS has calculated the width of $A$ by 
assuming $\tan\beta$ = 1 and $\cos(\beta -\alpha$) = - 0.06. 
In Fig.~\ref{fig:AZH_14}, we show the distribution of 
$\sigma \times Br(\rightarrow Zh \rightarrow  \ell \ell b \bar b)$ 
for all the points in the scanned data set, and then overlay 
the 95 $\%$ C.L. upper limits by the ATLAS (red solid line) 
and CMS (blue dashed line) collaborations 
\cite{ATL-PHYS-PUB-2013-016, CMS:2013dga} 
at 14 TeV LHC with 
$\lum = $ 3000~$\ifb$. We find that the ATLAS 
limits are more stronger (by almost one order) than the CMS, 
although the reason behind this is apparently not clear. From 
the Fig.~\ref{fig:AZH_14} it is clear that only a very small 
region of the parameter space will be excluded by the HL-LHC data, 
in fact to probe the remaining parameter space few orders of 
magnitude improvement in cross section measurement is required.

\subsubsection{ Search for H with di-higgs ($H\to hh\to b \bar b \gamma \gamma$) final states}
\label{sec:Htohh14tev}

In this subsection we discuss the possibility of 
observing the heavy CP even Higgs boson ($H$) at HL-LHC
with its decay to a pair of SM-like Higgses ($h$) with one Higgs decaying 
to $b \bar b$ and other to $\gamma \gamma$ modes. 
Single $H$ production cross section 
can be up to two orders of magnitude larger compared to 
the direct $h$ pair production cross section (see Table~1 of 
Ref.\cite{Bhattacherjee:2014bca})\footnote{The SM 
Higgs pair production cross section at next to leading order is 
about 34 {\rm fb} at 14 TeV LHC for $M_{h}=125$ GeV \cite{higgs_cs}.} 
depending on the choice of model parameters and it 
can also have non-trivial effects on the self coupling 
measurement of the 125 GeV Higgs \cite{Bhattacherjee:2014bca}. 
%Here, we follow the methodology of Ref.~\cite{Bhattacherjee:2014bca} 
%in order to perform a detailed collider analysis.  

In the MSSM, the production cross section of $H$ and its 
decay to a pair of SM-like Higgses crucially depends on 
the SUSY parameter space, mainly $M_A$ and $\tan\beta$. 
More precisely, below the $t\bar t$ threshold (350~GeV), 
the decay rate $\Gamma(H \to h h$) is substantial only 
for smaller values of $\tan\beta$. The most dominant 
production mechanism of $H$ is the ggF process 
although, for large or moderate 
$\tan\beta$, the bottom quark annihilation to $H$ 
($b \bar b \rightarrow H$) cross section can be substantial \cite{Han:2013sga}.
The final state 
signature, depending on the decay of the 
SM-like Higgses, includes, for example, $b \bar b b \bar b$, 
$b \bar{b} \tau^+ \tau^-$, $b\bar b W^+W^-$, $b \bar b \gamma \gamma$ etc.  
Among all these possibilities, $b \bar b b \bar b$ final 
state has the largest cross section. However, due to 
enormous QCD background this is one of the most challenging 
scenarios to be observed at the LHC\footnote{ It has been shown that the 
jet substructure technique can be very useful to separate the 
signal events from the large QCD and electroweak 
backgrounds \cite{deLima:2014dta}.}. 
On the other hand, even 
though the branching ratio for $b \bar b \gamma \gamma$ channel 
is very small (about 0.27 \%), it is the most promising channel due to 
large photon identification efficiency and very good resolution 
in the photon energy measurement. Hence using the di-photon invariant 
mass distribution one can easily reconstruct the Higgs boson mass 
and at the same time separate the signal from the SM 
background\footnote{In this case, the dominant backgrounds are 
$t\bar{t}h$ and the direct Higgs pair production ($hh$). 
Note that, the production cross section for both the Higgs pair 
and $t \bar t h$ processes depend on the MSSM parameters, however, here 
we assume SM cross sections for $hh$ and $t \bar t h$ processes.}.  

%--------------------------------------------------
\begin{figure}[!htb]
 \begin{center}
{\includegraphics[angle =0, width=0.6\textwidth]{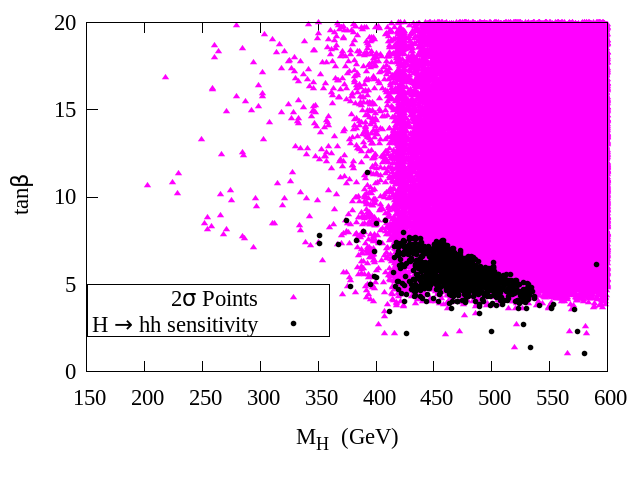}}
\caption{  {\it Sensitivity of di-Higgs final state in $M_A - tan\beta$ 
plane at HL-LHC 
from $b \bar b \gamma \gamma$ channel with $\lum = $3000~$\ifb$. The 
magenta coloured points are 2$\sigma$ allowed points 
from global analysis and the black circled points are expected 
to be probed at HL-LHC (see Sec.\ref{sec:Htohh14tev} for details).  
}}
\label{fig:Htohh}
 \end{center}
\end{figure}
%--------------------------------------------------

A detailed signal-background analysis of heavy Higgs 
production and its decay to a pair of 125 GeV Higgs 
in $b\bar b \gamma\gamma$ channel has been already 
performed in Ref.~\cite{Bhattacherjee:2014bca}. Here we use 
their results to constrain our parameter 
space\footnote{The 
analysis for $b \bar b \gamma \gamma$ channel 
was first introduced in the Ref.~\cite{Baur:2003gp}}. 
Events with two $b$-jets, 
two photons and no isolated 
leptons are selected after imposing the basic 
selection cuts following the 
ATLAS collaboration \cite{ATLAS-collaboration:1484890}. 
%We consider that the leading (sub-leading jets) 
%$b$-jet has $p_T >$ 40 (25) GeV and $|\eta|<$ 2.5 and also 
%assume b-tagging efficiency $\epsilon_{b} = $ 80\%. Photon 
%with $p_T >$ 25 GeV, $|\eta|<$ 2.5 satisfying the photon 
%isolation criteria are selected with identification 
%efficiency $\sim$ 80\%. For photon isolation we demand the 
%scalar $p_T$ sum of all stable visible particles 
%within a cone of radius $\Delta R $= 0.4 around the photon 
%candidate should not exceed 4 GeV \cite{Aad:2010sp}. Possible 
%overlaps between the selected $b$-jets and photon candidates 
%are elliminated imposing the following choices of 
%$\Delta R$: $\Delta R(b,b)$, $\Delta R(b,\gamma)$, 
%$\Delta R(\gamma \gamma) >$ 0.4. We then calculate the 
%invariant mass of the $b \bar b$ and $\gamma\gamma$ system 
%and demand that the reconstructed masses satisfy the following 
%criteria: $50$ GeV$<M_{b\bar b} <130$ GeV, 
%$120$ GeV $<M_{\gamma\gamma} <130$ GeV. 
Using the four momentum information of the $b\bar b$ and $\gamma\gamma$ system, 
the invariant mass of the heavy Higgs boson can be reconstructed 
assuming $M_{b\bar b \gamma\gamma}$ is $M_{H} \pm 50 ~{\rm GeV}$.

Let us now estimate the sensitivity of the 
HL-LHC to probe the parameter space in the 
$b \bar b \gamma \gamma$ channel. In 
Ref.~\cite{Bhattacherjee:2014bca}, authors 
considered few representative benchmark 
points with heavy Higgs mass $M_H$ in the 
range of 275 - 600 GeV and performed a detailed 
collider analysis. We should note, as also 
pointed out in Ref.~\cite{Bhattacherjee:2014bca}, 
the cut efficiencies do not change radically 
in the entire region of 275 - 600 GeV. The 
number of background events ($N_B$) and 
the cut efficiencies as a function of $M_H$ are taken from 
Ref.~\cite{Bhattacherjee:2014bca}. We estimate the 
number of signal events ($N_S$) for all the points in the  
scanned data set by multiplying the cut efficiencies 
with the production cross sections, and then 
calculate the signal significance ${\mathcal S} = N_{S} / \sqrt{N_B}$. 
In Fig.~\ref{fig:Htohh}, we present the sensitivity of the 
$b \bar b \gamma \gamma$ channel at HL-LHC where magenta triangles 
corresponds to our scanned data set, while the black circles 
represent the points with ${\mathcal S} > 2\sigma$. In other words,  
these are the points (black circle) with low $\tan\beta$ ($<$ 10) 
that are expected to be probed at the HL-LHC. The lack of sensitivity 
of HL-LHC in the regions of parameter space with $\tan\beta > 10$ 
is due to the fact that both the ggF production cross section 
and Br($H\to hh$) decreases with the increase of $\tan\beta$ forcing 
this portion of parameter space to go beyond the reach of HL-LHC. 
From the Fig.~\ref{fig:Htohh}, one can note that 
future run of LHC might not be able to exclude the entire 
region below $M_{H} \lsim $ 425 GeV and $\tan\beta \lsim $ 8.

\subsubsection{Search for $H/A$ with $t \bar t$ final state}

Above the kinematic threshold of $t \bar t$ ($\sim$ 350 GeV), the decay 
of $H$ and $A$ into $t \bar t$ pair opens up. In fact, for 
$M_{A} >$ 350 GeV and low to moderate values of $\tan\beta$, it 
is indeed the dominant decay mode of $H/A$. Besides, due to 
large top-Higgs Yukawa coupling, the production 
cross section of the heavy Higgs $H$ via ggF process 
also becomes large in the low $\tan\beta$ regime. Hence, 
assuming the narrow width approximation, one can expect 
to observe a resonance peak at $M_{H/A}$ in 
the $t\bar t$ invariant mass distribution. However, the 
main drawback of such a bump hunting procedure is that it 
is extremely difficult to extract the $t\bar t$ 
resonance peak from the huge SM $t\bar t$ continuum 
background. Here, we perform a detailed signal-background 
analysis in the context of HL-LHC and study the sensitivity 
of HL-LHC to probe the region of parameter space of 
our interest.

%--------------------------------------------------
\begin{figure}[!htb]
 \begin{center}
\includegraphics[angle =270, width=0.6\textwidth]{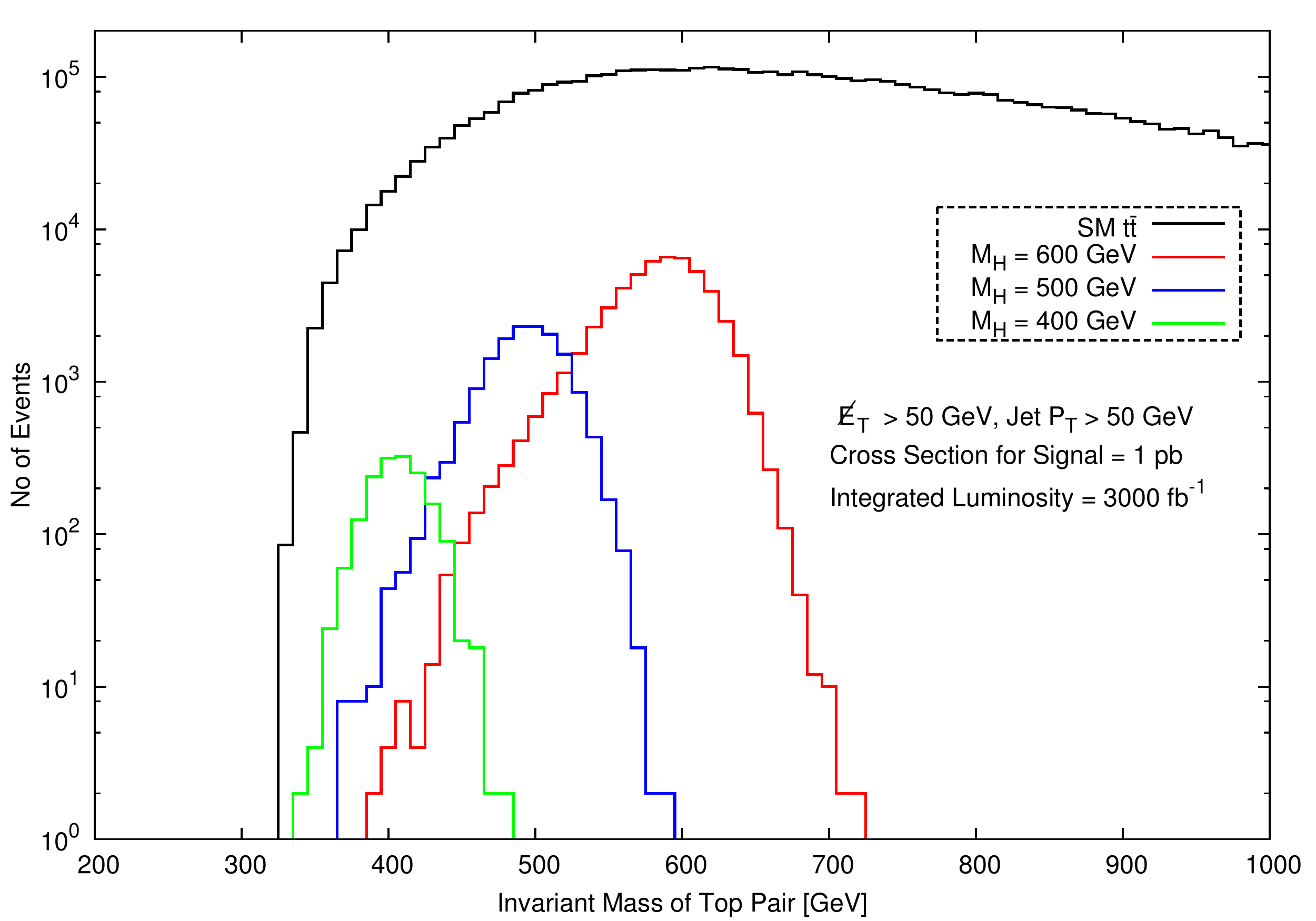}
\caption{ {\it Distribution of $t\bar t$ invariant mass for some 
representative SUSY benchmark points with $M_H$= 
400 GeV (green), 500 GeV (blue), 600 GeV (red) and 
SM $t\bar t$ background (black). The distribution has been drawn before 
applying the cut on invariant mass ($C5$) in signal region \it{SR-loose} 
(see text for details) for integrated luminosity $\lum$ = 3000~$\ifb$. 
We consider NNLO+NNLL cross section for SM $t\bar t$ production at 14 TeV LHC (966 pb)
% https://twiki.cern.ch/twiki/bin/view/LHCPhysics/TtbarNNLO
and for signal the no of events has been estimated assuming 
single $H$ production cross section to be 1 $pb$ with $Br(H \rightarrow t \bar t)$ = 100\%. 
}}
\label{fig:invmass}
 \end{center}
\end{figure}
%--------------------------------------------------

We analyze the production of heavy Higgses $H/A$ via ggF process 
and their decay to $t \bar t $\footnote{As $H$ and $A$ are almost 
degenerate, one might not be able to distinguish them in $t\bar t$ 
invariant mass distribution. For this reason, here we consider 
the production of the heavy Higgses $H/A$ simultaneously via 
the ggF process.}, where one top decays leptonically and one decays 
hadronically. So, the final state includes one isolated lepton (electron or muon), 
at least four jets among them at least two are 
$b$-jets and missing transverse energy ($\MET$). We use PYTHIA 
(version 6.4.28) \cite{Sjostrand:2006za} to generate both the signal and dominant 
SM background, e.g., $t\bar{t}$ events. Electrons are selected with $p_T >$ 20 GeV 
and $|\eta| <$ 2.47, while we choose muons with 
$p_T >$ 20 GeV and $|\eta| <$ 2.4. We select jets 
with $p_T >$ 30 GeV and $|\eta| <$ 2.8. A jet is called 
a ``b-jet" if the angular separation 
$\Delta R = \sqrt{(\Delta \eta)^2 + (\Delta \phi)^2}$ in the $\eta - \phi$ 
plane between the jet and a B-hadron is less than 0.2. Following 
the ATLAS collaboration, we also assume a flat 70\% b-tagging 
efficiency for the b-jets \cite{btagging}. In our analysis, we also 
implement the methodology of 
lepton isolation and lepton-jet identification 
following the ATLAS study \cite{Aad:2014wea}. Although it is possible 
to reconstruct leptonically decaying top quark 
within a quadratic ambiguity, for simplicity we use the 
parton level four momentum information of the neutrino 
to reconstruct the $H$ mass. 
%In our analysis, neutrino is the 
%main source of missing energy and z-component of neutrino momentum 
%can be calculated from the four momentum mass equation 
%$M_W^2 =(p_l + p_{\nu})^2$.        

We are now in a position to describe the details of 
our simulation procedure as well as the kinematic
selection cuts for our signal and the backgrounds. We 
choose three representative benchmark points with 
$M_{H}$ = 400, 500 and 600 GeV, and define three 
signal regions (SR), namely {\it{SR-loose, SR-medium, SR-tight}}, 
depending on the choice of our selection cuts. The SR-loose 
is defined based on the following set of cuts: 
\bi
\item C1: Events must contain at least 4 jets with $p_T >$ 50 GeV.
\item C2: Among the four selected jets, two are $b-tagged$. 
\item C3: Events containing one isolated lepton with $p_T >$ 30 GeV are selected. 
\item C4: Missing energy ($\met) >$ 50 GeV.  
\item C5: Select events with $t \bar t$ invariant mass between $M_{H} \pm $ 25 GeV. 
\ei
We denote the signal region {\it{SR-medium}} with the same set of cuts 
as that of {\it{SR-loose}} but with $\met >$ 100 GeV. In order to probe 
higher values of $M_H$, we define the signal region {\it{SR-tight}} where 
the cuts on $p_T$ of the jets and $\met$ are stronger. For example, for 
{\it{SR-tight}}, events are selected with $p_T$ of the first 
two leading jets greater than 100 GeV while the missing transverse 
energy $\met >$ 100 GeV.

%-------------------------------ttbar-------------------
\begin{table}[htb!]
\begin{center}
\begin{tabular}{|c |c|c|  c|c| c|c|}
\hline
Channel & \multicolumn{6}{c|}{Number of Events at 3000~$\ifb$} \\
\cline{2-7}
        & \multicolumn{2}{c|}{$M_H = 400$ GeV}&\multicolumn{2}{c|}{$M_H = 500$ GeV} & \multicolumn{2}{c|} {$M_H = 600$ GeV} \\
\cline{2-7}
                & Signal        & $t \bar t$    & Signal        &$t \bar t$     &Signal         &$t \bar t$             \\
\hline
$SR-loose$      & 1268          & 104612        & 9658          &420572         &26842          &563452         \\
\hline
$SR-medium$     &8              &1741           &1584           &69232          &9656           &194698         \\
\hline
$SR-tight$      &-              &-              &4              &637            &2296           &44894          \\
\hline
\end{tabular}
  \caption{\small \it \label{tab7}
Number of signal (H production via ggF) and SM $t \bar t$ events after 
applying the cuts 
in SR-Loose, SR-Medium and SR-Tight signal region at 14 TeV LHC with
$\lum$ = 3000~$\ifb$. For $t \bar t$, we use NNLO cross section \cite{ttbar_cs}.
For the signal estimation we present the number assuming
$\sigma(pp \rightarrow H)_{NLO} \times Br(H \rightarrow t \bar t)$ = 1 pb.
  }
\end{center}
\end{table}
%--------------------------------------------------

In Fig. \ref{fig:invmass}, we display the $t\bar t$ invariant 
mass distribution for the three representative benchmark points 
with $M_H$= 400 GeV (green), 500 GeV (blue), 600 GeV (red), and 
also overlay the same for the SM $t \bar t$ background (black line).  
We show the invariant mass distribution for the signal 
region {\it SR-loose} with an integrated $\lum$ = 3000~$\ifb$. To 
estimate the number of background events, we use the 
$t\bar t$ cross section at next-to next-to leading order (NNLO) 
$\sigma^{t \bar t}_{NNLO}$ = 966 pb \cite{ttbar_cs}. 
For simplicity, we assume that for all the benchmark points 
$Br(H \to t \bar t)$ is 100\% and the next-to leading order 
production cross section $\sigma(pp \rightarrow H)$ is 1 pb. 
Note that, we make such a conservative choice 
just to keep our analysis simple, one can easily scale 
our numbers with the actual production cross sections and branching 
ratios. Although one can see clear resonance peaks at the $M_H$ 
masses (see Fig.~\ref{fig:invmass}), enormous SM $t\bar t$ 
background makes it very challenging to observe a clear signal 
of heavy Higgs in the $t\bar t$ invariant mass 
distribution over the SM background. To be more precise, we 
count the number of signal and background events for 
three signal regions {\it SR-loose, SR-medium and SR-tight} after 
C5, and then calculate the statistical significance 
${\mathcal S} = N_{S}/\sqrt{N_B}$. In Table~\ref{tab7}, we present 
the number of signal and background events for three 
benchmark points and the SM $t\bar t$ backgrounds at the 
14 TeV run of LHC with $\lum$ = 3000~$\ifb$. From 
Table~\ref{tab7} one can see that 
$N_{S}/N_{B}$ ratio is very small for all 
the benchmark points irrespective of the signal regions.  
We find that the statistical significances ${\mathcal S}$ 
for $M_H$ = 600 GeV are 36, 22 and 11 for the 
three signal regions {\it SR-loose}, {\it SR-medium} 
and {\it SR-tight} respectively. However, even with 
5\% systematic uncertainty these numbers reduces 
to 0.95, 0.99, 1.02 respectively.

%--------------------------------------------------
\begin{figure}[!htb]
 \begin{center}
\includegraphics[angle =0, width=0.6\textwidth]{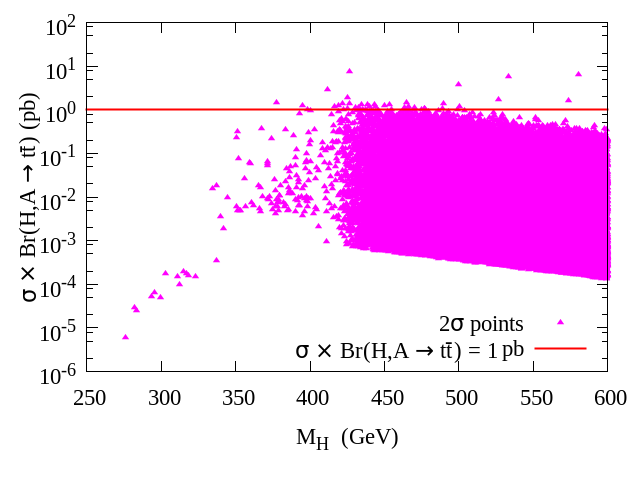}
\caption{ {\it
Scatter plot in
$M_{H} - [\sigma \times {\rm Br (H,A \to t \bar t)}]$
plane assuming both H and A production via ggF at 14 TeV LHC.
Magenta colored points represents for 2$\sigma$ allowed
parameter space. Red line corresponds to  $\sigma \times  Br(H,A \to t \bar t)$ =  1 pb.
}}
\label{fig:Htt14}
 \end{center}
\end{figure}
%--------------------------------------------------

In Fig.~\ref{fig:Htt14} we show the distribution of 
$\sigma \times  Br(H,A \to t \bar t)$ assuming production of 
both H and A via ggF for all the points 
in our scanned data set. We find that 
$\sigma \times  Br(H,A \to t \bar t)$ lies mostly 
in the region 0.5 - 0.0001 pb. The red solid line in 
Fig.~\ref{fig:Htt14} represents 
$\sigma \times  Br(H,A \to t \bar t)$ = 1 pb. This red solid 
line indicates that for most the points in our 
scanned data set the quantity cross section 
times $Br(H,A \to t \bar t)$ is 10 - 1000 times smaller, and 
thus the the numbers in Table~\ref{tab7} represent a too 
much optimistic scenario. In other words, even 
HL-LHC might not be sensitive enough to probe 
such a region of parameter space.

Recently, it has been shown 
that using angular cuts, the signal significance 
can be improved \cite{Djouadi:2015jea}. However, one should note that
the inclusion of systematic uncertainties may change the significance 
drastically and it is not possible to predict the reach of the 
$t\bar t$ channel at the HL-LHC without the precise knowledge 
of the systematic uncertainties. We end this section by 
mentioning that the observation of a heavy Higgs in 
$H \to t\bar t$ channel at the HL-LHC is really a 
challenging task and it needs special attention and 
more detailed studies.
%Here we would like 
%point that analysis in Ref.~\cite{Djouadi:2015jea} do not consider  
%systematic uncertainties while calculating the signal 
%significances, and we have just shown the significances 
%reduces drastically when one consider the role 
%of systematic uncertainty. 

\subsubsection{Search for $H/{A}$ with $\tau^{+}\tau^{-}$ final states}
\label{sec:Htautau_14}

In Sec.\ref{sec:Htautau_8} we study the impact of the 
$H/{A} \to \tau^{+}\tau^{-}$ channel on the parameter 
space of interest using the LHC-8 data 
and find that regions with large $\tan\beta$ ($>$20) 
and $M_A$ up to 600 GeV are already excluded. 
In this section, we discuss the sensitivity of the 
14 TeV run of LHC to probe the remaining allowed region 
of parameter space via the $\tau^{+}\tau^{-}$ channel. In 
order to perform a detailed signal-background analysis, we 
follow the ATLAS simulation with LHC-8 data \cite{Aad:2014vgg}. 
We assume that $H/A$ are produced via the b-quark 
associated production process and decays to $\tau^{+}\tau^{-}$. 
The ATLAS analysis \cite{Aad:2014vgg} have shown that for 
$M_{A}$ between 300 and 600 GeV, the final state with 
one leptonically decaying $\tau$ ($\tau_{\rm lep}$) 
and one hadronically decaying $\tau$ ($\tau_{\rm had}$) 
gives the best sensitivity (see Fig.~9(b) of Ref.~\cite{Aad:2014vgg}), 
and so in our analysis we 
consider the $H/{A} \to \tau_{\rm lep}\tau_{\rm had}$ signature 
only. We identify taus through their hadronic 
decays and demand that the candidate jet must lie 
within $|\eta|<$ 2.5 with $p_{T} >$ 30 GeV. The 
jet must contain one or three charged tracks with 
$|\eta_{\rm track}| < $ 2.5 and highest track 
$p_{T} > $ 3 GeV. Moreover, in order to ensure 
proper charge track isolation, we also 
require that there are no other charged 
tracks with $p_{T} > $ 1 GeV inside the 
candidate jet.

We select events with at least one lepton (electron/muon) with 
$p_{T} >$ 50~GeV and an oppositely charged $\tau$-hadron 
with $p_{T} >$ 50~GeV. We further demand that events do not 
contain any additional electrons or muons, and the 
transverse momentum difference 
$\Delta p_{T} \equiv p_{T}(\tau_{\rm had}) - p_{T}({\rm lepton})$ 
is greater 
than 50 GeV. Additionally, following Ref.~\cite{Aad:2014vgg}, we also 
demand the sum of the azimuthal 
angles $\sum \Delta \phi \equiv \Delta \phi(\tau_{\rm had},\MET) + 
\Delta\phi(\tau_{\rm lep},\MET)$ is less than 3.3, and the 
hadronic and leptonic $\tau$-decays satisfy 
$\Delta\phi(\tau_{\rm lep},\tau_{\rm had}) >$ 2.4. Besides, events 
with at least one b-tagged jet with $p_{T} >$ 50 GeV are also 
selected. Finally, we 
demand that the reconstructed di-tau invariant mass ($m_{\tau\tau}$) is 
within $M_{\Phi} \pm $ 30 GeV ($\Phi$ = $H/A$). The dominant 
SM background processes that can mimic the signal are $\rm Z$ +jets, $\rm W$ +jets, 
QCD multi-jets and $t\bar t$. We analyze the most dominant SM backgrounds, 
namely $\rm Z$ +jets and $t\bar t$, while we check that QCD 
multi-jet contribution is negligibly small. In order to accommodate other 
subdominant backgrounds, we consider 50\% enhancement of the 
total estimated background.

%--------------------------------------------------
\begin{figure}[!htb]
 \begin{center}
{\includegraphics[angle =0, width=0.48\textwidth]{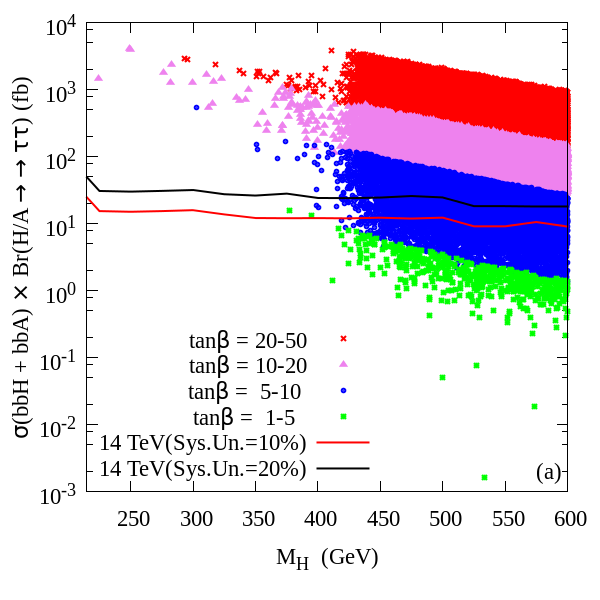}}
{\includegraphics[angle =0, width=0.48\textwidth]{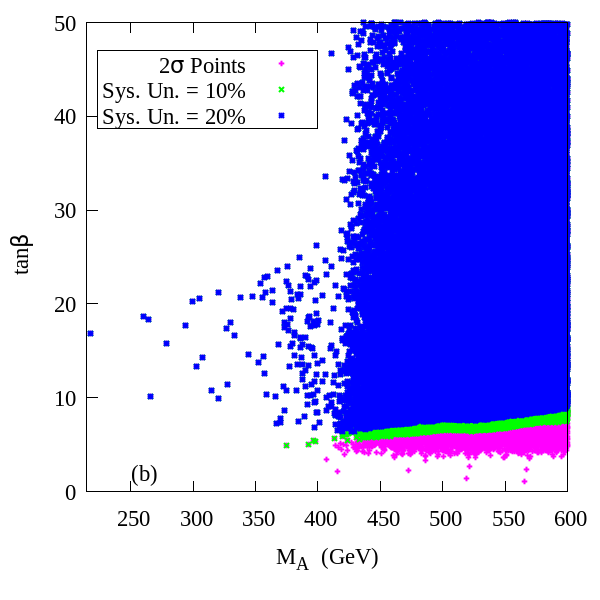}}
\caption{{\it Left: Scatter plot in 
$M_{H} - [\sigma \times {\rm Br (\Phi \to \tau^+ \tau^-)}]$ plane 
where $\Phi =$ H, A produced in association with $b$-quarks. The 
entire $\tan\beta$ region is splitted into different pieces, 
the regions with $20 < \tan\beta < 50$, $10 < \tan\beta < 20$,
$5 < \tan\beta < 10$ and $1 < \tan\beta < 5$
are shown in red (cross), pink (triangle), blue (circle)
and green (square) respectively. The solid red (black) 
line represents the expected upper limits 
on $\sigma \times {\rm Br (\Phi \to \tau^{+}\tau^{-})}$ at the 
HL-LHC with $\lum = $3000~$\ifb$. Right: Sensitivity 
of $H/{A} \to \tau^{+} \tau^{-}$ final states in 
$M_A - tan\beta$ plane. The magenta (plus) 
points are the 2$\sigma$ allowed points corresponding 
to our scanned data set, while the blue (square) and green 
(cross) points are expected to be probed at HL-LHC 
considering 20\% and 10\% systematic uncertainties 
respectively.}}
\label{fig:Htautau_14}
 \end{center}
\end{figure}
%--------------------------------------------------

The signal events are generated using PYTHIA (version 6.4.28) \cite{Sjostrand:2006za} 
assuming $H/A$ are produced via the b-quark associated production 
process. Here we concentrate on the region 
$200 < M_{A} < 600$ GeV and split the entire $M_A$ 
region in steps of 25 GeV. For each event we then calculate the 
di-tau invariant mass using the ``collinear 
approximation technique" with the assumption that the $\tau$-lepton 
and all its decay products are collinear \cite{Elagin:2010aw} and 
$\MET$ in an event is 
entirely due to the neutrinos\footnote{Another technique often used 
by the experimental collaborations are the Missing Mass 
Calculator (MMC) method which allows a better reconstruction 
of the $\tau\tau$ invariant mass distribution \cite{Elagin:2010aw}. 
In this paper, however, we restrict ourselves to 
the ``collinear approximation technique". }. After performing 
a bin-wise signal/background analysis, we thus obtain a 95\% C.L. 
expected exclusion limit on the production cross sections 
for the signal events 
assuming ${\rm BR}(H/{A} \to \tau^{+}\tau^{-})$ = 100\%. 
We calculate the production cross sections for all the points 
corresponding to our scanned data set, and then impose the 
estimated exclusion limits on the production cross sections 
for different values of $M_A$ (or $M_H$). In 
Fig.~\ref{fig:Htautau_14}(a), we display the distribution 
of the quantity 
$\sigma \times Br(H/{A} \to \tau^{+}\tau^{-})$ when $H/A$ 
are produced via the b-associated process. Similar to 
Fig.~\ref{fig:Htautau}, we here also display the scanned data set 
with different $\tan\beta$ dependences. For example, the 
regions with $20 < \tan\beta < 50$, $10 < \tan\beta < 20$, 
$5 < \tan\beta < 10$ and $1 < \tan\beta < 5$ 
are shown in red (cross), pink (triangle), blue (circle)
and green (square) respectively. The solid red and 
black lines indicate the expected sensitivity in the HL-LHC 
with 10\% and 20\% systematic uncertainties respectively. 
%% Note that, the selection cut efficiencies that we obtain for the signal events for three different values 
%% of $M_A$, namely $M_A$ = 300 GeV, $M_A$ = 400 GeV and $M_A$ = 500 GeV are 0.74, 0.68 and 0.54 respectively. Thus, 
%% for a given benchmark point, one can calculate the selection cut efficiencies for the background events and then using the 
%% above-mentioned signal efficiencies can estimate the possibility of 
%% probing the selected benchmark point via the $\tau^{+}\tau^{-}$ channel at the future high luminosity run of LHC.}

We have already mentioned that in the MSSM, the 
production cross sections and branching ratios of $H$ 
and/or $A$ crucially depend on two important parameters 
$M_A$ and $\tan\beta$. Thus the expected sensitivity 
of the 14 TeV LHC on the $H/A$ production 
cross sections can be translated in the 
$M_{A} - \tan\beta$ plane. In 
Fig.~\ref{fig:Htautau_14}(b) we display such possibility 
where the magenta (plus) points denote our scanned 
data set while the blue (squared) and green 
(crossed) points represent the 
regions of parameter space in the $M_{A} - \tan\beta$ plane 
which can be probed at the high luminosity run of LHC 
with 3000 $\rm fb^{-1}$ of data assuming 20\% and 10\% 
systematic uncertainties respectively.

Thus, combining the existing 8 TeV direct search 
bounds (see Fig.\ref{fig:Htautau}) and the estimated 
HL-LHC bounds, one can infer that the regions 
with $\tan\beta$ $>$ 20 are already excluded by 
LHC-8 data while the regions with 
$\tan\beta$ down to 8 with any values of $M_A$ 
can be probed at the HL-LHC.

%%--------------------------------------------------
%\begin{table}[htb!]
%\begin{center}
%\begin{tabular}{|c |c|c|  c|c| c|c|}
%\hline
%After cut /Cuts& \multicolumn{6}{c|}{Number of Events at 3000~$\ifb$} \\
%\cline{2-7}
%        & \multicolumn{2}{c|}{$M_H = 400$ GeV}&\multicolumn{2}{c|}{$M_H = 500$ GeV} & \multicolumn{2}{c|} {$M_H = 600$ GeV} \\
%\cline{2-7}
%		& Signal        & SM/dominant Z    & Signal        &SM/dominant Z    &Signal         &SM/dominant Z            \\
%\hline
%CUT....     	& 	        & 		   & 	      	   &		     &		     &				\\
%\hline
%CUT....     	& 	        & 		   & 	      	   &		     &		     &				\\
%\hline
%CUT....	      	& 	        & 		   & 	      	   &		     &		     &				\\
%\hline
%Upper limit on  & \multicolumn{2}{c|}{ }&\multicolumn{2}{c|}{ } & \multicolumn{2}{c|} { } \\      
%$\sigma_{BSM} (95\%~C.L.)$ 		& \multicolumn{2}{c|}{--- fb}&\multicolumn{2}{c|}{ ---fb} & \multicolumn{2}{c|} {----fb} \\
%\hline
%\end{tabular}
%  \caption{\small \it \label{tab8} ----------- Cut efficiencies and Upper limit on new physics cross-sections at 14 TeV LHC }
%\end{center}
%\end{table}
%%--------------------------------------------------
%

%-----------------
%----------------------------------
\vskip 0.4cm
\section{Conclusion }
\label{sec5}

A scalar particle with mass close to 125 GeV 
has been discovered at the LHC. Measurements of spin-parity 
and various couplings seem to agree with the SM 
expectations. In the minimal extension of SM, namely MSSM, 
one can identify the observed 125 GeV Higgs boson as 
the lightest Higgs boson among the five MSSM Higgses 
$h$, $H$, $A$ and $H^{\pm}$. In the MSSM, 
the couplings of $h$ with the SM gauge bosons ($W/Z$) 
are proportional to $\sin(\beta-\alpha)$ where 
$\tan\beta$ is the ratio of vevs of two Higgs doublets while 
$\alpha$ is the Higgs mixing angle. Precise measurements of 
various couplings of the observed Higgs boson with $W/Z$ bosons by the 
ATLAS and CMS collaborations imply $\sin(\beta-\alpha) \sim 1$. 
At the tree level, the Higgs mixing angle can be derived using 
pseudoscalar Higgs mass parameter $M_A$ and 
$\tan\beta$, however if we include radiative 
corrections, $\alpha$ becomes a non-trivial function of various 
SUSY parameters. The ATLAS and CMS collaborations have 
also searched for the heavy Higgs bosons ($H$, $A$, $H^{\pm}$), 
however absence of any signal puts strong bounds on the 
masses/branchings of these heavy Higgses. One can easily 
satisfy the current LHC data by taking 
$\alpha \rightarrow 0$, and $\beta \rightarrow \pi/2$ with 
$M_A >> M_Z$, which is generally known as the decoupling limit of MSSM. 
In this limit, the lightest MSSM Higgs boson behaves exactly like 
SM Higgs and masses of other Higgs bosons ($H$, $A$, $H^{\pm}$) 
are pushed well above LHC reach. In this paper we study 
the possibility of having light additional MSSM Higgs 
bosons, preferably below 600 GeV, with moderate 
Higgs mixing angle $\alpha$, being consistent with the 
SM Higgs data and also direct search limits on the MSSM 
heavy Higgses.

We restrict ourselves to the 19 dimensional pMSSM framework 
and scan the parameters that are relevant for the MSSM 
Higgs sector. We perform a global fit analysis 
using most updated data (till December 2014) from 
the LHC and Tevatron experiments and also consider 
the flavor physics constraints. The region with 
$\rm {M_{A} \le 350 ~GeV}$
and $\rm {\tan\beta} \ge 25$ are excluded by the 
$\rm {Br (B_{s} \to \mu^+ \mu^-)}$ while 
$\rm {M_{A} \le 350 ~GeV }$ with $\rm {\tan\beta} \le ~8$ is 
not favoured by the $\rm {Br (b \to s \gamma)}$ constraint. 
The regions with large $M_A$ value ($>400$ GeV) are 
not much constrained by the data. An interesting point 
to note that regions with $200<M_A<400$ can have moderate 
values of the Higgs mixing angle $\alpha$ = 0.2, while 
for relatively large values of $M_A$, $\alpha$ can be 
as large as $\sim$ 0.8 with small $\tan\beta$. Thus, one 
is not always forced to be in the decoupling limit 
to comply with the LHC data and light additional Higgses ($M_A$) are 
still allowed by the current data. Moreover, 10 - 20\% 
deviations from the SM expectations are also observed for 
various Higgs signal strength variables.

We next study the impact of current bounds on the MSSM heavy 
Higgs boson ($H$, $A$ and $H^{\pm}$) masses and 
couplings from the direct search at the LHC. 
We analyze the following decay modes of the MSSM 
heavy Higgses: $H \to \gamma \gamma$, 
$H \to WW$, $H \to hh \to b \bar b b \bar b $, $H \to hh \to b \bar b \gamma \gamma$, 
$H/A \to \tau^+ \tau^-$, $A \to Zh$, $H^\pm \to \tau^\pm \nu$ and $H^+ \to t \bar b$. 
As we have already mentioned, most of 
the regions of parameter space satisfy 
the alignment limit ($\beta - \alpha \sim \pi/2$), the 
7+8 TeV LHC data on $H \to WW$ channel is 
not sensitive enough to constrain the 
parameter space of our interest. Except $H/A \to \tau^+ \tau^-$, 
LHC bounds on the production cross section times branching ratios 
for other decay modes of $H$, $A$ and $H^{\pm}$ are 
also one/two orders of magnitude larger compared to the MSSM 
expectations. The 7+8 TeV LHC data on MSSM heavy Higgs boson decay 
$\Phi (=H/A) \to \tau^+ \tau^-$ put the most stringent bound on our 
parameter space. The entire region with $\tan\beta >$ 20 is excluded 
when the heavy Higgses $\Phi = H/A$ are produced via $b\bar b \Phi$ 
process. The reason being large $\tan\beta$ (typically $>20$) implies large 
Br($H/A \to \tau^+ \tau^-$), typically $\sim 10\%$, and large 
cross section for associated production of $H/A$ with bottom 
quarks, and thus stronger constraint on the quantity 
$\sigma \times Br$.

We also study the prospect of probing the allowed 
parameter space, mostly low $\tan\beta$ region, at the high 
luminosity run of LHC (HL-LHC). We impose the ATLAS and CMS 
preliminary results on $H \to ZZ$ and 
$A \to Zh$ at the HL-LHC. The impact of the 
future limits from the $H \to ZZ$ and 
$A \to Zh$ channels on the allowed 
parameter space are found to be marginal at 
$\lum$ = 3000~$\ifb$. 
Searches via 
$H \to hh$ at the HL-LHC 
can only probe a small part of the 
parameter space with low $\tan\beta$.  
Among several other possible decay modes, 
$H \to t \bar t$ dominates in the low $\tan\beta$ 
region for $M_{H} >$ 350 GeV. We 
perform a dedicated signal-background analysis on 
the $H \to t \bar t$ channel by choosing few representative MSSM 
benchmark points, and find that the signal to background 
ratio is very small. A more detailed analysis is required, for 
example one can use the jet substructure technique, spin 
correlation technique to achieve better sensitivity in this 
channel. We find that so far the combined 7+8 TeV LHC data 
on the $H/A \to \tau^{+}\tau^{-}$ channel has the best 
sensitivity to place constraints on the MSSM parameter 
space of interest. Thus, one may expect to find 
stronger constraints on the MSSM parameter space 
from the $\tau^{+}\tau^{-}$ mode at the HL-LHC. 
We study this possibility and find that 
combining the existing 8 TeV direct search bounds 
and the estimated HL-LHC bounds, one can infer that 
the regions with $\tan\beta$ $>$ 20 are already excluded 
by LHC-8 data while the
regions with $\tan\beta$ down to 8 with low to moderate 
values of $M_A$ can be probed at the HL-LHC. 
Below $\tan\beta < 8$, searches via $H \to hh$ at the HL-LHC 
can be very important to probe the parameter space.

In summary, we find that regions with 
low to moderate values of 
$\tan\beta$ with light additional Higgses 
(mass $\le$ 600 GeV) remain unconstrained by the 
current data. Even the high luminosity run of LHC may 
not have enough sensitivity to probe the entire 
low $\tan\beta$ 
region of parameter space. However, the 
proposed $e^{+}e^{-}$ international 
linear collider (ILC) will be an ideal 
machine to study this scenario. With 
the expected accuracies in the determination 
of various partial decay widths of the Higgs 
boson and also the possibility of producing 
some of the heavy MSSM Higgses directly, one 
might be able the probe the remaining region of the 
allowed parameter space with 
$\sqrt s = 1000$ GeV with higher luminosities 
at the ILC.

%----------------------------------
\vskip 0.4cm
%\bigskip
\small
\baselineskip 16pt
{\large \bf \underline {Acknowledgements:}}
\vskip 0.12cm
We acknowledge Shankha Banerjee for useful discussions regarding global fits. 
Work of B. Bhattacherjee is supported by Department of Science and Technology, Government
of INDIA under the Grant Agreement numbers IFA13-PH-75 (INSPIRE Faculty Award). 
A. Chakraborty would like to thank the Department of Atomic Energy, 
Government of India for financial support. The work of A. Choudhury 
was partially supported by funding available 
from the Department of Atomic Energy, Government of India, for 
the Regional Centre for Accelerator-based
Particle Physics (RECAPP), Harish-Chandra Research Institute.

%----------------------------------
%\bibliographystyle{unsrt}

%----------------------------------

\end{document}

%%%%%%%%%%%%%%%%%%%%%%%%%%%%%%%%%%%%%%%%%%%%%%%